NOBEL SYMPOSIA

# Microfluidics

## Nobel Symposium 162

*Microfluidics: Proceedings of the 162nd Nobel Symposium,*
Stockholm, Sweden, 5-8 June 2017
Ed. Johan Elf, Irmeli Barkefors and Thomas Lurell

# Nobelsymposium 162 – Microfluidics

*June 5-8, 2017*

*Sång-Säby Conference Center, Stockholm*

## Introduction

On June 5, 2017, the long awaited date for the Nobel Symposium in Microfluidics at Sånga Säby conference resort in the countryside west of Stockholm had arrived. Full of expectations, the hosts for the symposium, Johan Elf and Thomas Laurell, welcomed attendees to an exciting four days Odyssee through the discoveries and developments within the microfluidics and lab-on-a-chip field.

The Nobel Foundation's Symposium programme was initiated in 1965 and since then over 160 symposia have taken place. With the Nobel symposium nr 162, we had the ambition to cover the past, current and future developments of microfluidics; a highly challenging task as there are many excellent researchers in this area and only thirty slots in the program.

We tried to cover the early breakthrough contributions to the field, important developments over the years and applications of microfluidics that now propagate in vastly different directions both as industrial components or processes and as tools and methods supporting fundamental research.

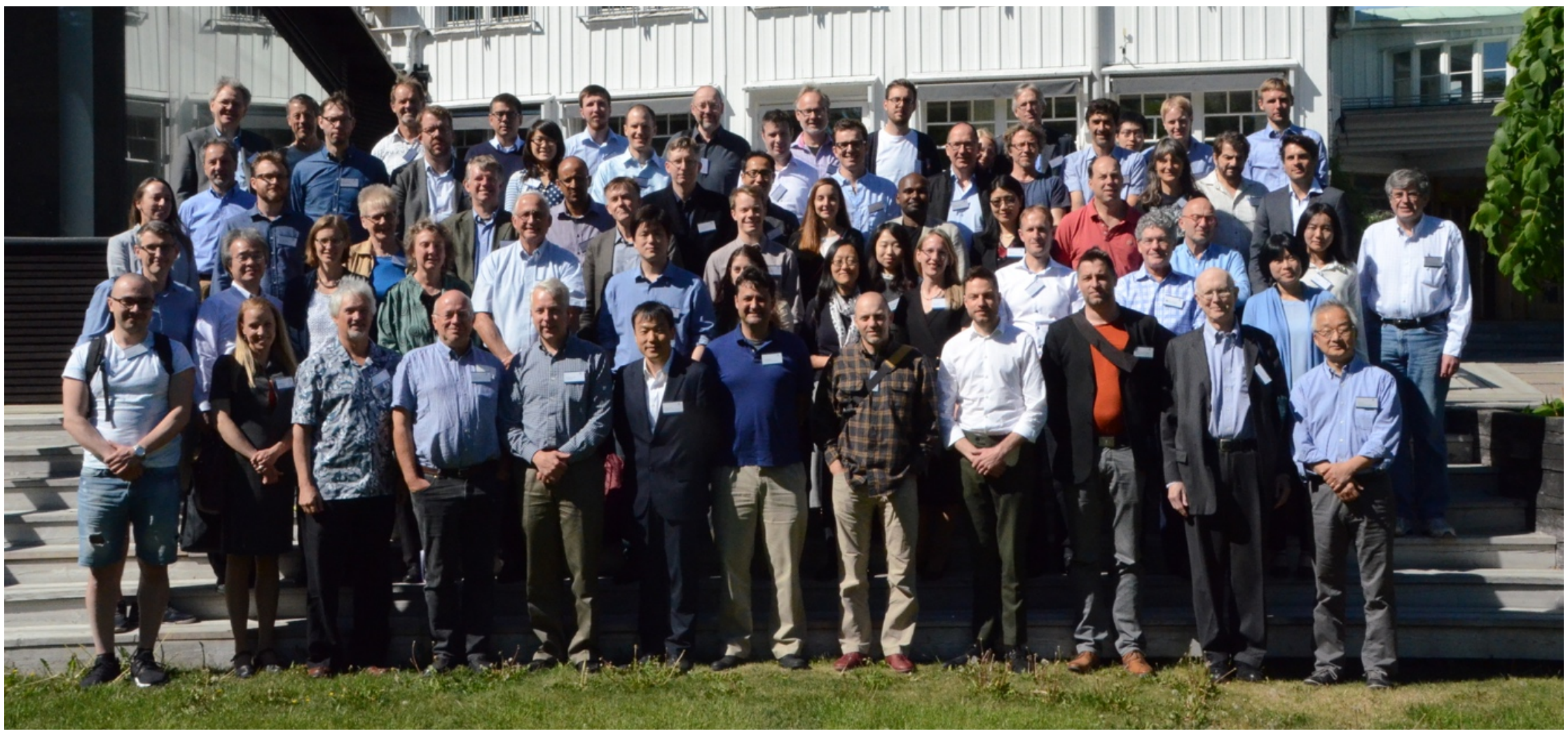

*Attendees of the No 162 Nobel Symposium – Microfluidics lined up in the sun in front of the sundeck at Sånga Säby conference resort.*

The opening lecture of the symposium was given by Professor George Whitesides who set the stage and the level for the week to come with statements that challenged the auditorium in terms of questioning why we do what we do and how the research performed can benefit society both in terms of creating industries, facilitate health care and benefit mankind.

The symposium program was mixed with some of the early pioneers in the field featuring Andreas Manz, Jed Harrison, Michael Ramsey and Richard Mathies, who all reflected on the early days of microchip based separations and the boom and hype created in the era of CE driven genome sequencing. A large part of the symposium highlighted the different development routes that the microfluidics field subsequently followed as material and fabrication technologies matured, enabling simple manufacturing of microfluidic circuits without the need for clean room facilities.

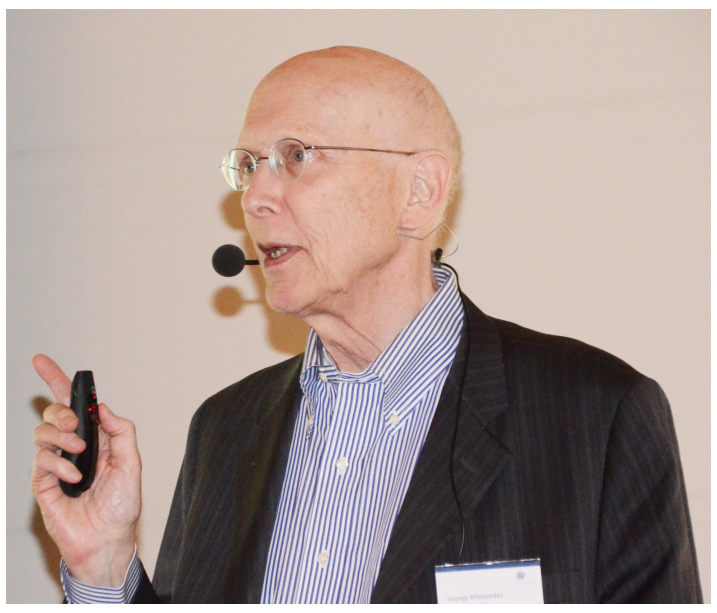

*George Whitesides, Stockholm, Sweden, 2017*

Important contributions and application developments were highlighted by Takehiko Kitamori, James Landers, Klavs Jensen, Albert van den Berg, Luke Lee and not least David Weitz, who took us through the decades describing a research branch that went from science fiction, through the stages of infinite promise, brisk reality check and finally scientific maturity and commercial and industrial implementation. That the field will continue to thrive also in the future was made clear by Sindy Tang, Amy Herr, Paul Blainey, Tomas Knowles, Andrew DeMello, Hang Lu, Yoon-Kyoung Cho and Roland Zengerle who convinced the audience that the only real limits are those of your imagination and that indeed microfluidics can solve real problems.

Mehmet Toner demonstrated that microfluidics has a place also in the macro-world, showing the ability to scale microfluidics to clinical applications still benefitting from the high performance of microscale. David Bebee followed the “make it simple” mantra and demonstrated important applications of autonomous microfluidic systems and simple solutions to clinical applications. Shoji Takeuchi outlined new strategies to address the unmet needs in regenerative medicine and Jay Han’s pulled the attention to microfluidic opportunities in chemical engineering for molecular and ionic separations.

That the microfluidic techniques are increasingly used as workhorses in biological research was obvious from the presentations by Stephen Quake, Jianhua Qin, Sabeth Verpoorte, Petra Dittrich and Joel Voldman, while Henrik Bruus presented theoretical aspects of microscale acoustofluidics and novel acoustophysics only revealed in confined space.

Having had the opportunity to gather this extraordinary line up of internationally top researchers in the field of microfluidics was a once in a life time event and indeed the euphoria was mutually contagious between the participants, sparking lively discussions late at night and throughout the week, with full attendance until the very end. No one wanted to miss anything! In summary, we as symposium chairs (Johan and Thomas) were privileged to get the opportunity to organize Nobel Symposium No. 162, devoted to the topic of Microfluidics.

NOBEL SYMPOSIA

To give the entire research community an opportunity to sample the science presented at the symposium we have compiled a symposium volume that reflects the major messages conveyed by our invited speakers. We are deeply grateful to the effort that our participants put into writing their lecture notes/abstracts. Most notably we all acknowledge the efforts that Irmeli Barkefors put into all the practical arrangements and herding of the participants throughout the week.

*Thomas Laurell & Johan Elf*

## The Nobel Symposia

The Nobel Foundation's Symposium programme was initiated in 1965. Since then about 160 symposia have taken place. The Nobel Symposia programme has obtained a good reputation and now plays a significant role in the international exchange of knowledge and experience.

The financial base for the Nobel Symposia programme consists of funds received from Riksbankens Jubileumsfond and the Knut and Alice Wallenberg Foundation.

## The Organizers

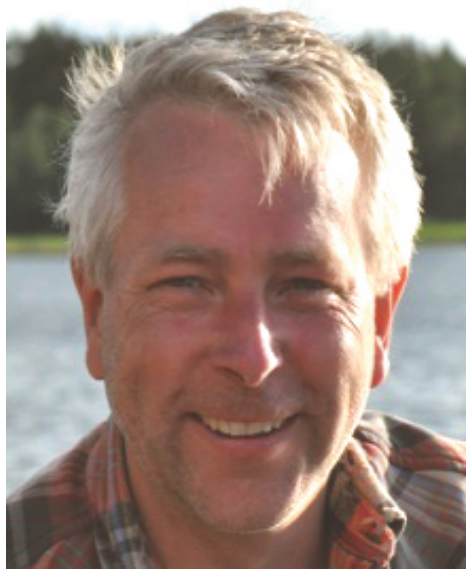

**Thomas Laurell** is Professor in Medical and Chemical Microsensors since 2000 with a focus on Lab-On-A-Chip technologies in biomedicine at the Department of Biomedical Engineering, Div. Nanobiotechnology and Lab-on-a-chip. He received his PhD in electrical engineering in 1995 at Lund University and obtained a position as associate professor in 1998 at Lund University performing research on lab-on-a-chip technology interfaced to mass spectrometry proteomics and disease biomarker research as well acoustic manipulation of cells and particles in microfluidics systems. In 2005 Laurell co-founded the Chemical and Biological Microsystems Society, CBMS, the ruling body of the MicroTAS conference series and served as the President of CBMS, 2009-2017. He has also cofounded the Centre of Excellence in Biological and Medical Mass Spectrometry, a national infrastructure node at Lund University.

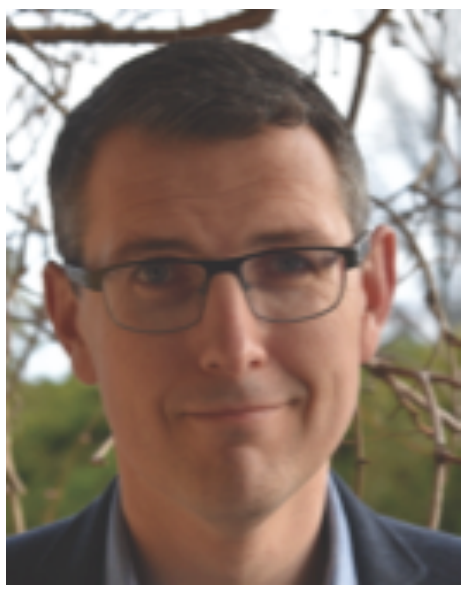

**Johan Elf** is Professor in Biological Physics at Uppsala University. Together with his interdisciplinary research team, Johan is pioneering single-molecule tracking methods by means of fluorescence microscopy in living cells. After obtaining his PhD on a theoretical subject in 2004, Johan Elf spent a more experimentally oriented postdoc period in the lab of Sunny Xie at Harvard University, developing the first methods for studying transcription factor binding at the level of single molecules in living cells. In the years following his return to Uppsala in 2007, Johan has focused his efforts on the study of gene regulation in bacteria, successfully combining theoretical and computational biology with experimental work and methods development to tackle challenging biological problems related to the central dogma of molecular biology.

# List of participants

## Speakers
Henrik Bruus (DTU)
Takehiko Kitamori (Tokyo)
David Weitz (Harvard)
Andreas Manz (KIST-Europe)
Sindy Tang (Stanford)
Shoji Takeuchi (Tokyo)
Hang Lu (GATech)
Luke Lee (Berkeley)
Jed Harrison (Alberta)
Stephen Quake (Stanford)
David Beebe (Wisconsin)
George Whitesides (Harvard)
Jay Han (MIT)
Andrew Demello (ETHZ)
Stefan Löfås (GE)
J. Michael Ramsey (North Carolina)
Richard Mathies (Berkeley)
James Landers (Virginia)
Amy Herr (Berkeley)
Joel Voldman (MIT)
Yoon-Kyoung Cho (UNIST)
Albert van den Berg (Twente)
Sabeth Verpoorte (RUG)
Klavs Jensen (MIT)
Petra Dittrich (ETHZ)
Jianhua Qin (DICP)
Mehmet Toner (Harvard/SBI)
Roland Zengerle (IMTEK)
Paul Blainey (Broad)
Tuomas Knowles (Cambridge)

## Observers
Sara Linse (LU)
Gunnar von Heijne (SU)
Heiner Linke (LTH)
Johan Åqvist (UU)
Olof Ramström (KTH)
Cleas Gustafsson (GU)
Mats Larsson (SU)
Jonas Tegenfeldt (LTH)
Wouter van den Wijngaart (KTH)
Fredrik Höök (Chalmers)
Aman Russom (KTH)
Johan Nilsson (LTH)
Per Augustsson (LTH)
Klas Hjort (UU)
Maria Tenje (UU)
Ove Öhman (Ginolis/Astrego)
Fredrik Westerlund (Chalmers)
Helene Andersson Svahn (KTH)
Håkan Jönsson (KTH)
Astrid Gräslund (SU)

## Organizers
Johan Elf (UU)
Thomas Laurell (LU)
Irmeli Barkefors (UU)

## Young researchers
Tatsuro Nakao (Tokyo)
Simon Berger (ETHZ)
Kethleen Bates (GATech)
Elisabet Rosas (Berkeley)
Isaac Micheal (UNIST)
Mathieu Odijk (Twente)
Dominik Hummer (ETHZ)
Soohong Kim (Broad)
Xiangping Li (KIST-Europe)
Hyun-Kyung Woo (UNIST)
Emre Iseri (KTH)
Michael Lawson (UU)
Özden Baltekin (UU/Astrego)
Pelle Ohlsson (LU)
Andreas Lenshof (LU)
Marc Isaksson (LU)
Carl Johanesson (LU)

# Practical (Simple) Microfluidics

*David J. Beebe - University of Wisconsin-Madison*

## Biography

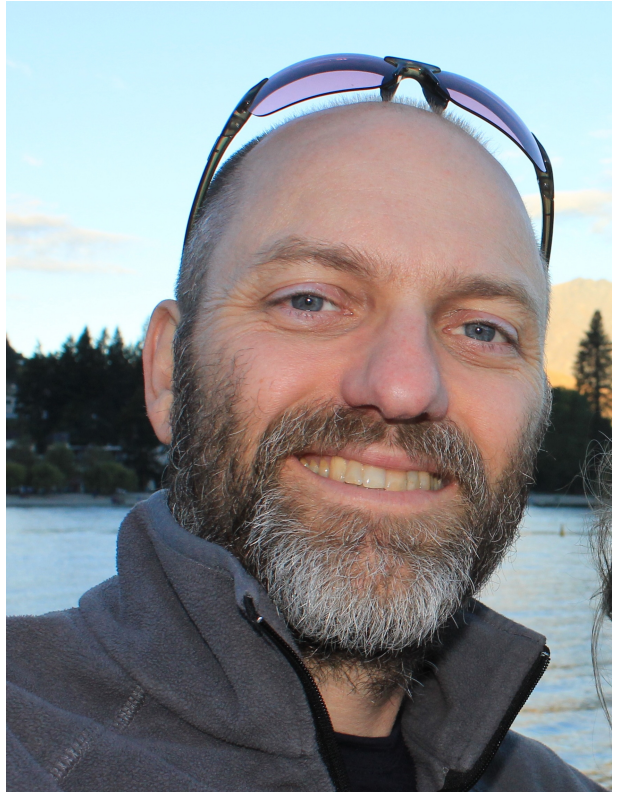

*David J. Beebe is the Claude Bernard Professor of Biomedical Engineering, Co-Leader of the Tumor Microenvironment Program in the UW Carbone Cancer Center, and a John D. MacArthur Professor at the University of Wisconsin-Madison. David's current research interests center on the understanding and application of micro scale physical phenomena to understand cancer biology (e.g. stromal-epithelial and cell-matrix interactions), to improve cancer diagnosis and monitoring and to advance global disease diagnostics.*

## Context

Back in the early 1990's I recall giving (and listening to) presentations having the obligatory list (a fairly long one) of all the potential advantages of this exciting new field of microfluidics. I won't dwell on the details, but suffice it to say that many of those supposed advantages either didn't turn out to be true or didn't turn out to be as sizable of an advantage as originally imagined. Despite the oversized expectations and never ending hope of a "killer app", significant progress has been made and microfluidics is having impact in a number of areas (albeit often hidden from view). Here, I'll focus on consistent themes that have been central to the path we have taken as we strive to do something useful with microfluidics. Our focus has always been less on the engineering and more on the physics. That is to say we strive to find simple ways to leverage the dominance of certain phenomena at the micro scale to create *simple*, hopefully *practical* and sometimes *fun* functionality.

## Influences

While we are all influenced by all that came before, I was probably most influenced by the work of two physicists - Edward Purcell and Robert Austin. Purcell illustrated the importance of understanding the basic physical phenomena at play and the elegance with which organisms have adapted to and leveraged such phenomena (Purcell, 1976). Austin expanded this notion by illustrating how to exploit this understanding of the physics in clever and fun ways to create intriguing micro scale devices and systems capable of new things. By way of example they also showed that one doesn't need to be bound by disciplinary borders.

## Simple equations

While I like to say I've focused on the physics, I'm not a physicist. So our way of "focusing on the physics" is by way of basic equations of transport (Fick's Laws) and interfaces (Laplace's Law) as well as dimensionless numbers prevalent in engineering textbooks (I believe there are over 50 such numbers). These simple equations allow one to think intuitively (for better or worse) without getting dragged into the details of the physics (apologies to the physicists - for a detailed examination of the physics of these numbers in the context of microfluidics see review by Squires & Quake, 2005). While Reynolds number (inertial, viscous) dominated in the early days of the field, I'm particularly fond of Peclet (diffusion, convection), Capillary (viscous, capillary)(Atencia, 2005), and, more recently, Bond (or Eötvös) (surface tension, body) numbers (Berry, 2011).

## Simple Examples

I provide a few representative examples of our work here. Additional examples in presentation.

*Autonomous operation (Simple & Fun)*

One of our first forays into leveraging physics in microfludics was to take advantage of short distances to create autonomously operating systems. While stimuli responsive materials (specifically hydrogels) were well known and well known to operate very slowly, we simply shrunk them down and used them inside microchannels to create autonomously operating components (e.g. valves, Beebe, Nature, 2000; Yu, APL, 2001) and systems (e.g. closed loop feedback, Eddington, LOC, 2001).

*Open microfluidics (Simple, Practical & Fun)*

About the same time, we began to think about surface tension (i.e. Young-LaPlace equation & Bond number) and create "open" micro scale devices - broadly defined as having at least one "side" open to the environment (or lacking a solid wall). Our initial work focused on surface patterning to control fluid with only a top and a bottom surface in contact with the fluid and the sides having contact only with air forming "virtual" walls (Zhao, Science, 2001). We also developed passive pumping schemes based on the principles described by Young-LaPlace equation (Walker, LOC, 2002). More recently we have expanded these concepts into "suspended microfluidics (Casavant, PNAS, 2013).

## Where to next

*Applications to biology and medicine (Simple & Practical)*

Since retraining in cancer biology (2004-2009) we have turned much of our attention to the application of microfluidics to both study basic biology as well as improve clinical decision making. We have learned that very simple embodiments of microfluidics (e.g. compartmentalization) can prove quite useful in exploring important questions in cancer biology (e.g. fibroblast-epithelia interactions, Sung, Integrative Biology, 2011), examining multi-kingdom interactions (e.g. mammalian-fungi signaling, Berthier, Plos Pathogens, 2013), and predicting patient response (e.g. stromal-epithelia interactions, Pak, Integrative Biology, 2015). Leveraging surface tension can allow discovery of new molecular interactions (e.g. protein-protein binding, Moussavi-Harami, J Proteome Research, 2013), low cost diagnostics (e.g. HIV viral load, Berry, PlosOne, 2015) and multi-omic analysis from a single sample (e.g. Sperger, Clinical Cancer Research, 2016).

## Acknowledgements

From 1996-1999 I was at the University of Illinois Urbana Champaign (UIUC). At UIUC, Juan Santiago, Jeff Moore, Matt Wheeler, Mark Shannon, Richard Magin and Ron Adrian all influenced my thinking in important ways. At the University of Wisconsin I'm indebted to Anna Huttenlocher, Caroline Alexander and Josh Lang for helping me see beyond the physical science of things. And thanks to the many current biological and medical collaborators that are too numerous to list. And most importantly all the students, post-docs and scientists who have passed through the lab over the years - they are the heart and soul of the work.

# ‘Applied Nanofluidics’ for Biomedical and Chemical Engineering

*Jongyoon Han - Massachusetts Institute of Technology*

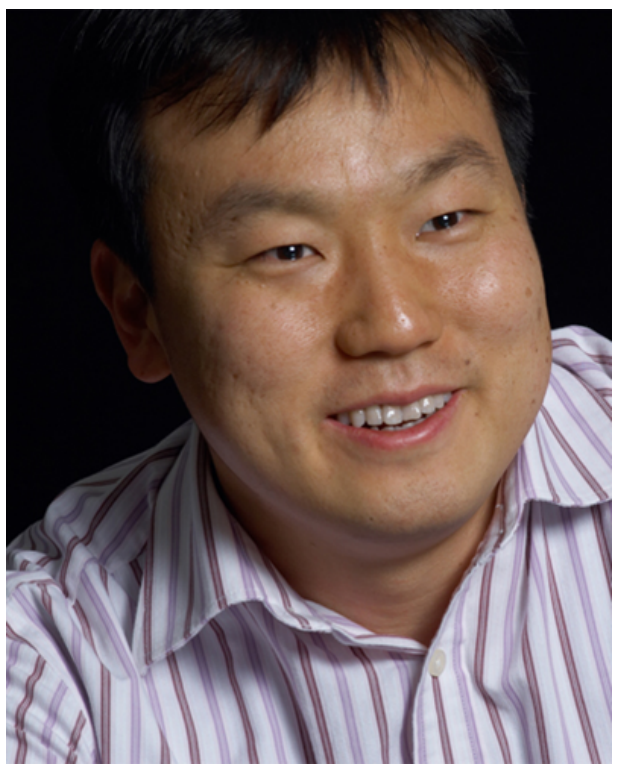

## Biography

*Prof. Jongyoon Han is currently a Professor of Electrical Engineering as well as a Professor of Biological Engineering, at the Massachusetts Institute of Technology. Dr. Han and his group pioneered the field of nanofluidic systems, such as entropic trapping DNA separation[1], efficient biomolecule concentration systems[2, 3], Anisotropic Nanofilter Array (ANA) structure[4] for continuous size-separation of proteins. Due to his significant contribution to nanofluidic device engineering, he received 2009 Analytical Chemistry Young Innovator Award from the American Chemical Society. He is also a recipient of NSF CAREER Award (2004), and Van Tassel Career Development Chair in Biomedical Engineering (2004). Originally focused at biomolecule separation and sample processing, his research has recently been making significant contributions in different application areas such as water purification[5], advanced neural prosthetics[6], and rare blood cell sorting[7], and mesenchymal stem cells (MSCs) engineering[8]. In addition, he is quite interested in leading international research and educational collaborations. He is currently participating in Singapore-MIT Alliance for Research and Technology Centre, remotely managing a thriving research group in Singapore. He is also a lead-PI for Kuwait-MIT Signature Project on Brine Desalination.*

## Abstract

Nanofluidic systems, with critical dimension of 10~100nm, are ideal model system for various types of nanoporous membranes. Artificially fabricated nanofluidic channels and nanopores have been successfully used for studying fundamental molecular / fluidic transport behaviors in nanoscale, yet there is much potential to take advantage of nanofluidic systems for real-world engineering applications. This seminar will showcase several recent examples of such applications my group was working on.

*Nanofluidic molecular separation and concentration*

Previously, we have demonstrated slanted nanofilter arrays and its implementation for continuous size-base protein concentration and separation. Using the nanofilter arrays, we have developed a system allowing to perform both purity and quality test for protein drugs, continuously. Not only could we perform multiple orthogonal analyses in a single device platform, but also the separation resolution and detection sensitivity match or exceed current tools used in the industry.

*Nanofluidics as a model ion selective membrane*

Ion concentration polarization is an ion transport process that occurs in the vicinity of ion selective membranes, which can be created and modeled by micro-nanofluidic systems. We have utilized this fundamental phenomenon for various applications, including biomolecule preconcentration, water purification / desalination, and even ion-specific nerve modulation. While the study of this multi-physics phenomenon has been challenging, experiments in nanofluidic systems allowed one to build a predictive, first-principle-based numerical models, with significant implications on membrane science and engineering.

# Electrophoretic cytometry for high-selectivity proteomics

*Amy E. Herr - University of California, Berkeley*

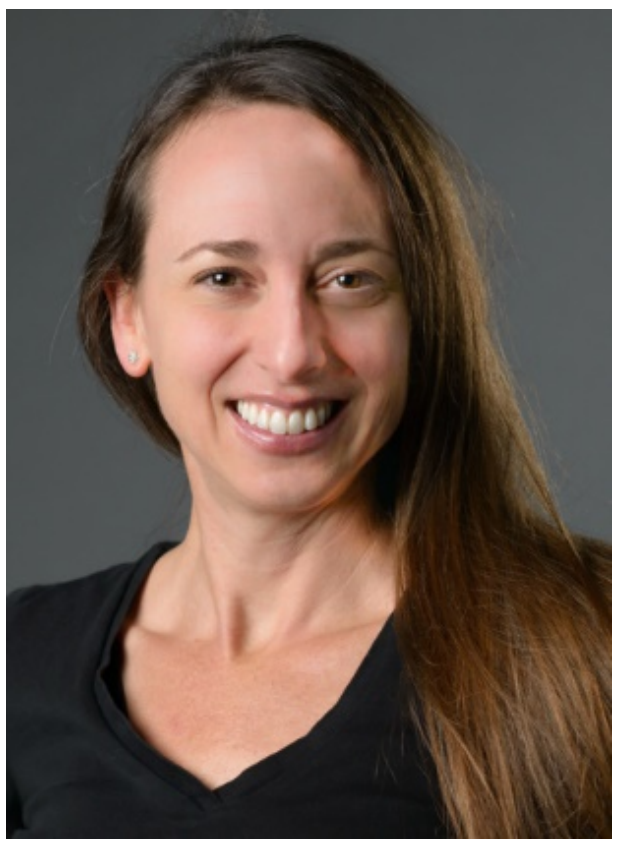

## Biography

*Amy E. Herr is the Lester John & Lynne Dewar Lloyd Distinguished Professor of Bioengineering at the University of California, Berkeley and a Chan Zuckerberg (CZ) Biohub Investigator. Prior to joining UC Berkeley, she was a staff member at Sandia National Laboratories (Livermore, CA), earned Ph.D. and M.S. degrees in Mechanical Engineering from Stanford University, and completed her B.S. in Engineering and Applied Science with honors from the California Institute of Technology. Her research has been recognized by the NIH New Innovator Award, NSF CAREER Award, Alfred P. Sloan Fellowship (Chemistry), and DARPA Young Faculty Award. Professor Herr has chaired the Gordon Research Conference (GRC) on the Physics & Chemistry of Microfluidics. She is an elected Fellow of the American Institute of Medical and Biological Engineering (AIMBE), an entrepreneur, and was recently elected to the US National Academy of Inventors. Her research program lies at the intersection of engineering design, analytical chemistry, and targeted proteomics – with a recent focus on cytometry spanning fundamental biological to clinical questions*

## Abstract

From fundamental biosciences to applied biomedicine, high dimensionality data is increasingly important. In cytometry tools, microfluidic design has played a key role in advancing the throughput, multiplexing and quantitation needed to generate rich data sets. Genomics and transcriptomics are leading examples. Yet, measurement of proteins lags. While proteins and their dynamic forms are the downstream effectors of function, the immunoassay remains the *de facto* standard (flow cytometry, mass cytometry, immunofluorescence) for direct measurement of endogenous, unmodified proteins. We posit that to realize the full potential of high-dimensionality cytometry, new approaches to targeted protein measurement are needed.

To address this gap, we have recently introduced a new class of 'electrophoretic cytometry' tools that increase target selectivity beyond simple immunoassays. Enhanced selectivity is essential for targets that lack high quality immunoreagents – as is the case for the vast majority of protein forms (proteoforms). In fundamental engineering and design, I will discuss how the physics and chemistry accessible in microsystems allows both the "scale-down" of electrophoresis to single cells and the "scale-up" to concurrent analyses of large numbers of cells. Particular emphasis will be placed on precision control of fluids and materials transport in passive systems, with no pumps or valves. Precise reagent control allows for integration of cytometry with sophisticated sample preparation – the unsung hero of measurement science.

Currently, our cytometry research extends two canonical targeted proteomic assays – isoelectric focusing and western blotting – to single-cell resolution. Single-cell isoelectric focusing (scIEF) is well-suited to profile protein post-translational modifications (PTMs) that drive intracellular signaling. Measurement of cell-to-cell variation in PTM-mediated signaling is challenging for two reasons. Namely, Immunoreagents lack the selectivity to distinguish the literally thousands of unique and combinatorial chemical modifications that are possible among proteoforms. This challenge plagues immunoassays. Compounding this is the fact that PTMs may change the molecular mass of a protein by a negligible mass difference, thus rendering protein sizing alone ineffective in resolving the resulting myriad, subtle proteoforms. To overcome these PTM-related challenges in cytometry, we describe isoelectric focusing followed by an immunoassay (i.e., immunoprobed IEF). While IEF is a mature separation mechanism, state-of-the-art IEF has not yet measured cell-to-cell variability owing to analytical sensitivity limitations [2]. Using precision control of mass transport and a novel 3D 'stackable' microfluidic device, we bring single-cell resolution to IEF.

Our scIEF devices are multi-layer microfluidic device. A 40-μm thick hydrogel "base" layer houses microwells that isolate individual cells. When seated on the base layer, a 500-μm thick hydrogel "lid" layer, patterned with three distinct chemistries – each containing the appropriate buffers and ampholytes – delivers lysis reagents and establishes the pH gradient necessary for IEF. After chemical lysis of cells (30 s), an electric field is applied and each single-cell lysate focuses (E=200 V/cm; 360 s). We observe that the hydrogel lid minimizes diffusive protein losses, yielding successful IEF on proteins from single cells. An arrayed format simultaneously assays multiple individual cells in parallel. We observe pH gradients that are linear, stable, and capable of resolving proteins differing by <0.1 pH units. scIEF is rapid, completing in <10-min. After focusing, benzophenone incorporated in the base layer is activated with UV (45-sec), thus covalently immobilizing the IEF-resolved proteins in the base gel layer. Finally, in-gel immunoprobing reports protein identity, correlated with isoelectric point.

The scIEF assay has reported down to single charge-unit differences in proteoforms from single cells engineered to express GFP. We see highly selective measurements of subtle chemical protein modifications cells as key to advancing a deeper understanding of the complex signaling networks that underlie stem cell differentiation and cancer. Our single-cell western blotting (scWB) assays are designed to resolve proteins based on molecular mass differences (electrophoretic mobility differences) by employing molecular sieving of cell lysate via polyacrylamide gel electrophoresis (PAGE). Similar in workflow to the scIEF assays, the scWB chips are a single-layer device that uses convective flow (pour over) to introduce lysis buffer and utilizes that buffer as the PAGE run buffer. We have demonstrated analyses of up to 15 proteins in a single cell, with lower detection limits estimated at ~27,000 protein copies. Bringing a multi-layer device architecture to the scWB has allowed us to adapt the assay to a compartment-specific, single-cell western blotting for nucleo-cytoplasmic profiling, which eliminates the need for complex image segmentation algorithms.

Looking forward, we will detail how these and other electrophoretic cytometry assays address driving cytology needs. We will consider ways in which high dimensionality data that includes targeted proteomic content enriches our understanding of health and disease. Here we will specifically consider the role of protein signaling and truncated isoforms in development of breast cancer drug resistance and the variable activation of protein signaling among even small populations of individual circulating tumor cells.

# Innovating Microfluidics and Pioneering Nanofluidics

*Takehiko Kitamori - The University of Tokyo*

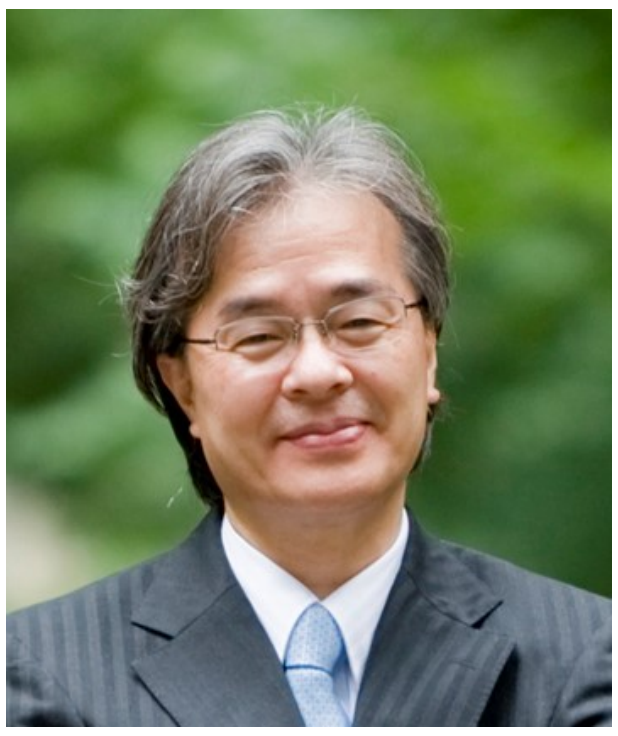

## Biography

*Professor Kitamori is a Professor in the Department of Applied Chemistry, School of Engineering, The Uni-versity of Tokyo. He was Dean of the Graduate School of Engineering (2010-2011), Vice President (2012-2013) and Special Assistant to the President (2014), and now he has returned to regular professor to devote himself to research and education. His extensive professional career includes positions at Hitachi and Kanagawa Academy of Science and Technology. He has received numerous honors for his innovative re-search, including an Award from the Science and Technology Agency, Chemical Society of Japan Award for Creative Work, and an Honorary Doctorate from Lund University, etc.*

## 1. Introduction

As well known, microfluidics was initiated by on-chip electrophoresis. Around the same time but a few years later, pressure driven microfluidics was opened independently. Our contribution to this field is to initiate the pressure driven microfluidics, and to pioneer nanofluidics. Regarding to microfluidics, we proposed basic concept and methods to innovate microfluidics for pressure driven and widely applicable device technology more than electrophoresis, and demonstrated some typical applications which have been still main current of microfluidics today.

Another contribution is to pioneer nanofluidics of which channel width is thousand folds smaller than microfluidics, while the concepts and methods are the same. Nanofluidics has been initiating ultimate small scale analytical and general chemistry evolving from mole to number of molecules. It surely brings breakthrough to other science and technology also, e.g., single cell single molecule analysis, absolute analysis by molecule counting, and solution chemistry and fluid mechanics at molecule level, which may largely contribute to life science and medicine.

Our social contributions to the research community are to found international and domestic academic societies, to launch new paper journals and international conferences. This lecture introduces these contributions and gives a view of prospects.

## 2. Microfluidics platform

Our contribution to both micro- and nanofluidics is establishing the basic platforms for general use of the pressure driven microfluidics. Microfluidics is now mainly going into applications. Immunoassay, cell culture, and extraction are the typical cases, and they need various kinds of chemical operations like mixing and reaction, extraction, phase separation, and so on. These chemical operations which are called as unit operation use not only aqueous solutions but organic solvents and various kinds of fluid control which are never achieved by electrophoretic molecule control and electroosmotic fluid control. Our group was the earliest one realizing these unit operations and fluid control in microfluidics and opening it to the variety of analytical, biological and even synthesis devices in the 1990's, and summarized as micro unit operation MUO in the early 2000's as shown in Fig. 1. The methods are as followings:

m-1) pressure and surface driven fluid control

m-2) MUO and their combination for complex multi step chemical processing

m-3) thermal lens micro and ultrasensitive detection for almost all species (non-fluorescent)

m-4) chemical functionalization of channel surface for above MUO and fluid control

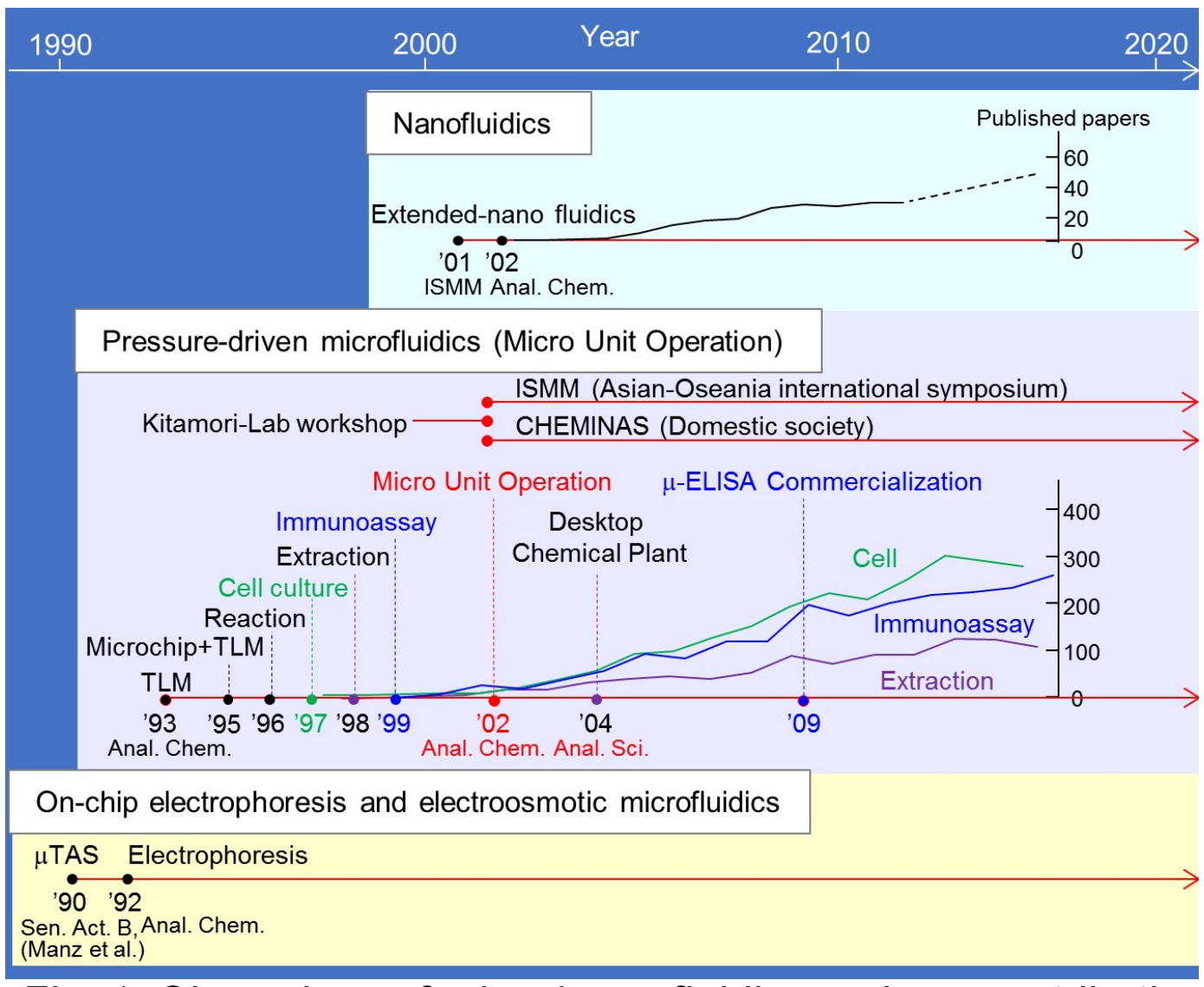

*Fig. 1. Chronology of micro/nano fluidics and our contribution.*

These four concepts and technologies harmonized to establish our methodology of the pressure driven microfluidics, and developed many kinds of applications because of its flexibility.

*Table 1. Micro Unit Operations (MUO).*

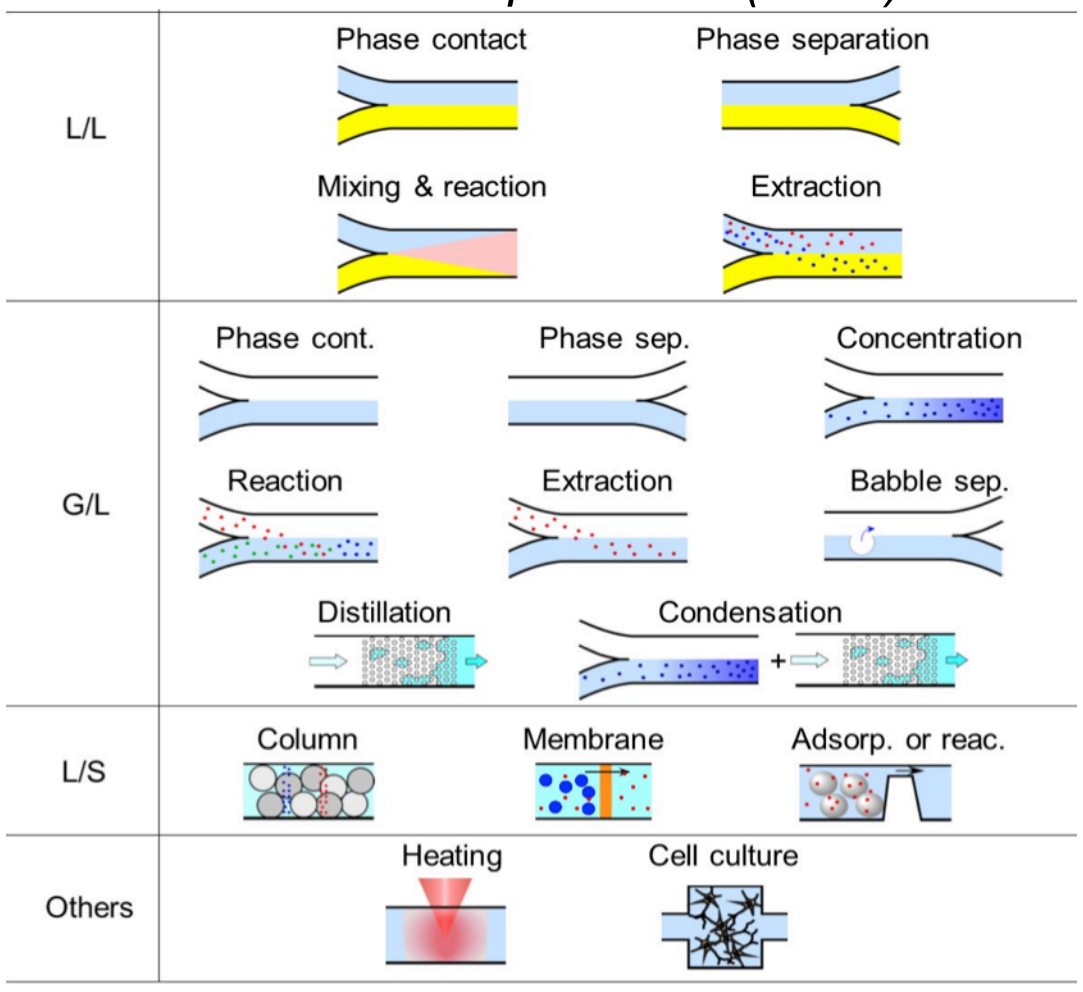

First of all, we developed our original detection method Thermal Lens Microscopy (TLM) in 1993, which is a kind of photothermal laser spectroscopy enabling detection of almost all molecules having optical absorption at single to countable number level in liquid. Its principle is shown and explained in Fig. 2a and its caption, and TLM is suitable for readout of target molecules in microchannel for general use. By the way, our first microchannel was fabricated on a slide glass to experiment TLM detection in liquid, because the space under the microscope was very narrow. This microchannel was Y-shaped channel

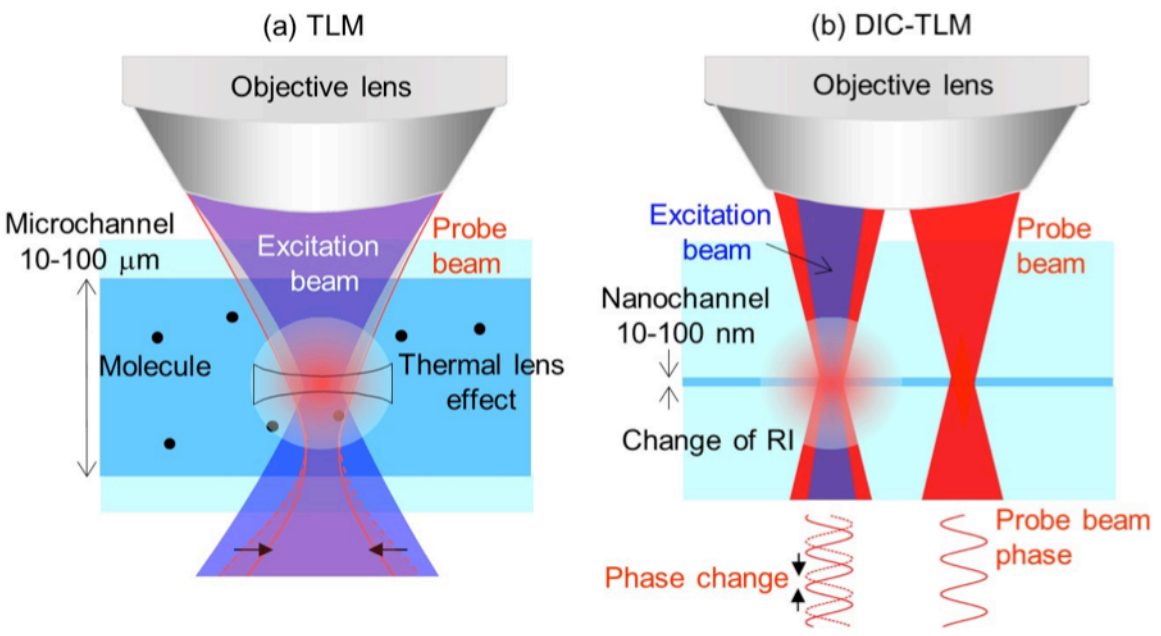

*Fig. 2. (a) Thermal lens microscope (TLM): Target molecules absorb the excitation light and emit heat via non-radiation relaxation. The emitted heat changes refractive index of the solvent around the focal area and its distribution acts as a concave lens which is called as the thermal lens effect. The probe beam detects the lensing effect which is in proportion to the number of molecules. Almost all molecules have this effect, and therefore, the TLM can detect almost all kind of the target molecules at countable number level. (b) Differential interference contrast thermal lens microscope (DIC-TLM): Nanochannel is smaller than the wavelength, and the thermal lensing effect which is a geographic optical effect is not valid. Therefore, the refractive index change induced by the target molecules is detected by optical interference between the two probe beams. DIC-TLM also can detect almost all kinds of the target molecules in nanochannels very high sensitively.*

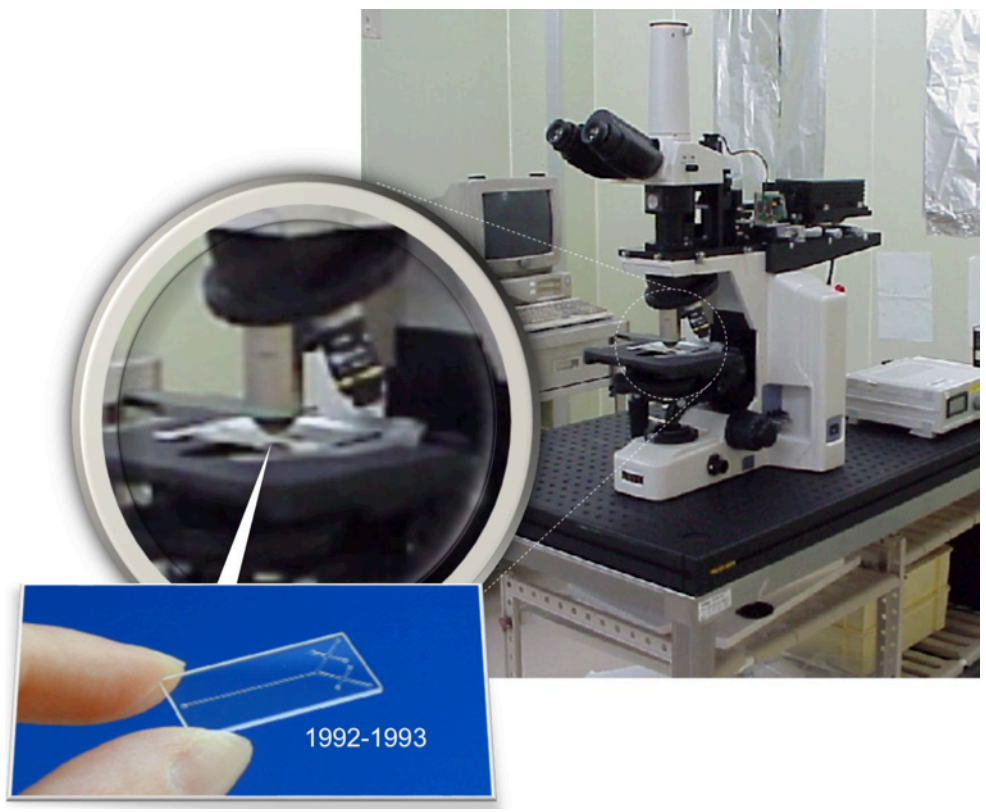

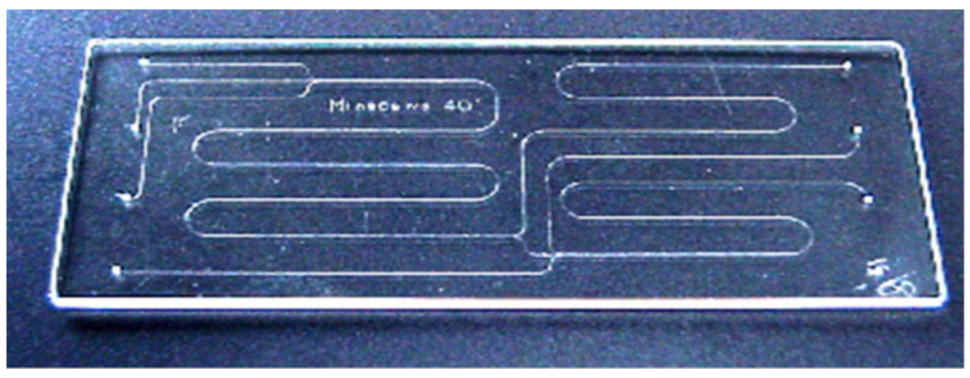

*(Top) Fig. 3. Desktop TLM, and our first microfluidic chip for developing the TLM. (Bottom) Fig. 4. A microfluidic chip for wet-analysis of metal pollutant in water. The target pollutant metal ion was chelated in the water sample, and the metal chelate was extracted to the organic solvent. Subsequently, the extraction phase of the metal chelate was purified by NaOH and HCl solutions, and was sensitively determined by the TLM.*

as shown in Fig. 3 and became the origin of MUO which are listed and explained in Table 1. Therefore, our strategic two methods of MUO and TLM were combined each other from 1995. This combination worked well to develop many kinds of chemical and biological processing, mixing reaction in 1996, cell culture in 1997, extraction in 1998, immunoassay in 1999, and summarized as MUO in 2002 as shown in Fig. 1. Figure 4 shows a microfluidic chip for analysis of a trace metal pollutant in water as a typical example. This microfluidics included 10 MUOs and used water sample, organic solvent, and several kinds of aqueous solution of reagents. This very complicated chemical process could be integrated on a pressure driven microfluidic chip, and the analysis time was reduced from several hours to a few minutes and LOD was improved 1000 folds by TLM detection comparing to the same analysis in ordinary macroscale. This device proved the excellent natures of microfluidics.

These applications have been main activities of microfluidics since 2000's, where hundreds of papers are published every year worldwide as shown in Fig.1, and today, they are actively developing to more sophisticated methodology regardless of droplet or continuous fluidics, and of glass or plastic substrate.

Space doesn't permit me to explain in details of these applications, but some of them have been put into practical use. For example, our microfluidic immunoassay system was commercialized as shown in Fig. 5 and could save some patients' lives in our University Hospital by utilizing its flexibility, rapidity and high sensitivity. A desktop chemical plant which contains thousands of microchannels for chemical production has been also realized by a company with our collaboration. On demand production by flexible operation, space saving by high efficient chemical processing, and other merits are expected.

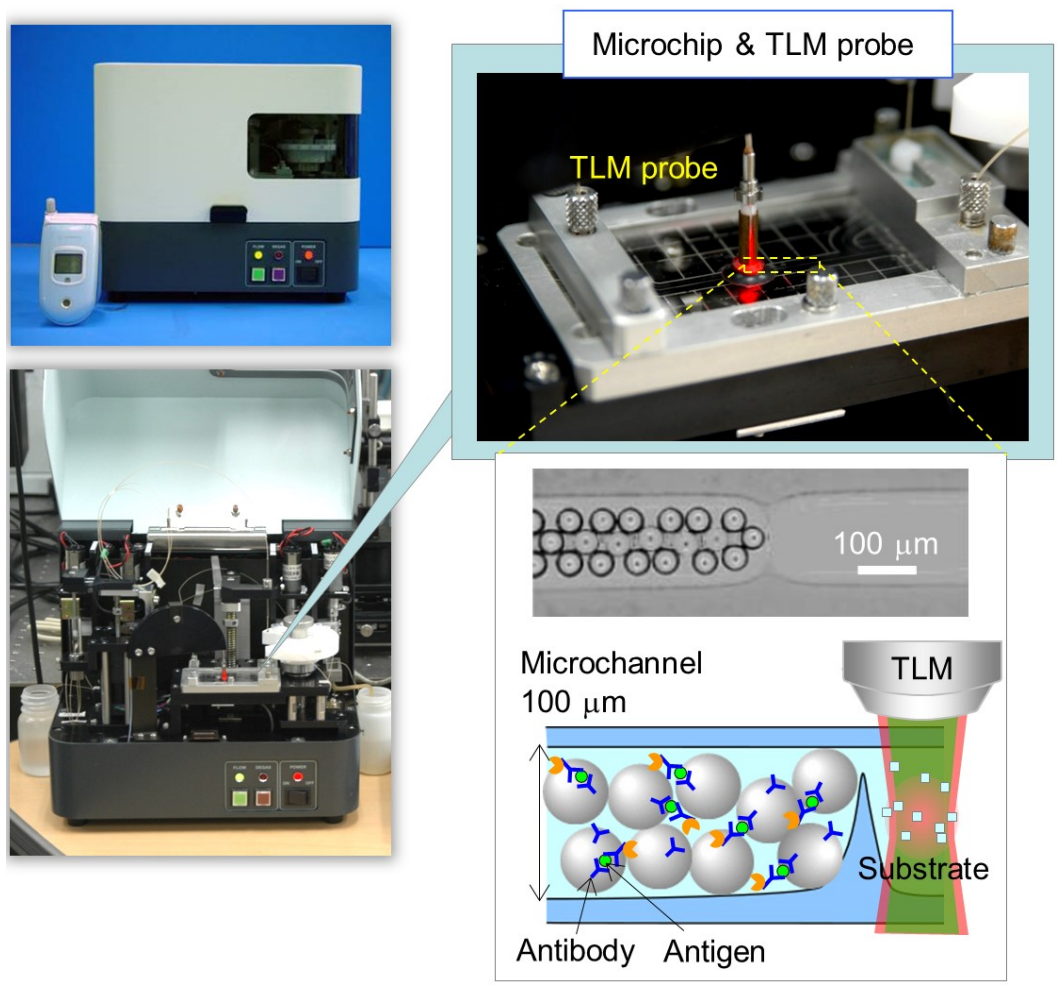

*Fig. 5. Commercialized microfluidics based enzyme-linked immunosorbent assay (ELISA) system, and its schematic illustration of the working principle.*

## 3. Nanofluidics platform

Another direction of progress in microfluidics is nanofluidics which is 1000 folds smaller in size than microfluidics and is smaller than wavelength of light. However, the principle of our methodology is basically the same as microfluidics listed above from m-1) to m-4);

n-1) pressure and surface driven fluidics,

n-2) nano scale unit operation,

n-3) photothermal readout of target molecules,

n-4) surface chemical functionalization,

though some modifications are needed for adjusting to nanoscale fluidics. We published a paper concerning the nanofluidic device in 2002 as shown in Fig. 1, and it was the earliest reports of nanofluidics using nanofabricated channels on a chip. Regarding readout, optical path length crossing nanochannels is too short for TLM, and therefore, we added wave optics to TLM for detect localized photothermal effect from the target molecules as described in the caption of Fig. 2b. After that, we have devoted ourselves to the extended-nano fluidic device technology and intensively reported fabrication and detection technologies.

Nano fabrication techniques of the Si semiconductor were modified for the fused silica and glass substrates which are just Si oxide and hard enough for nano fabrication and structures. The breakthrough technologies from micro to nano fluidics are,

bn-1) partial surface modification,

bn-2) low temperature glass bonding.

Inside of a nanochannel is quite unique space influenced by surface of the channel wall. Therefore, chemical surface control is much more definitive than a microchannel, and partial surface modifications at carefully designed local spots, bn-1), are needed. Therefore, our low temperature bonding, bn-2), between a cover glass substrate and a fabricated and surface functionalized glass substrates is quite important to keep the surface chemical functions inside the nanochannels. There are other techniques for nanofluidics, but they are left out for want of a space.

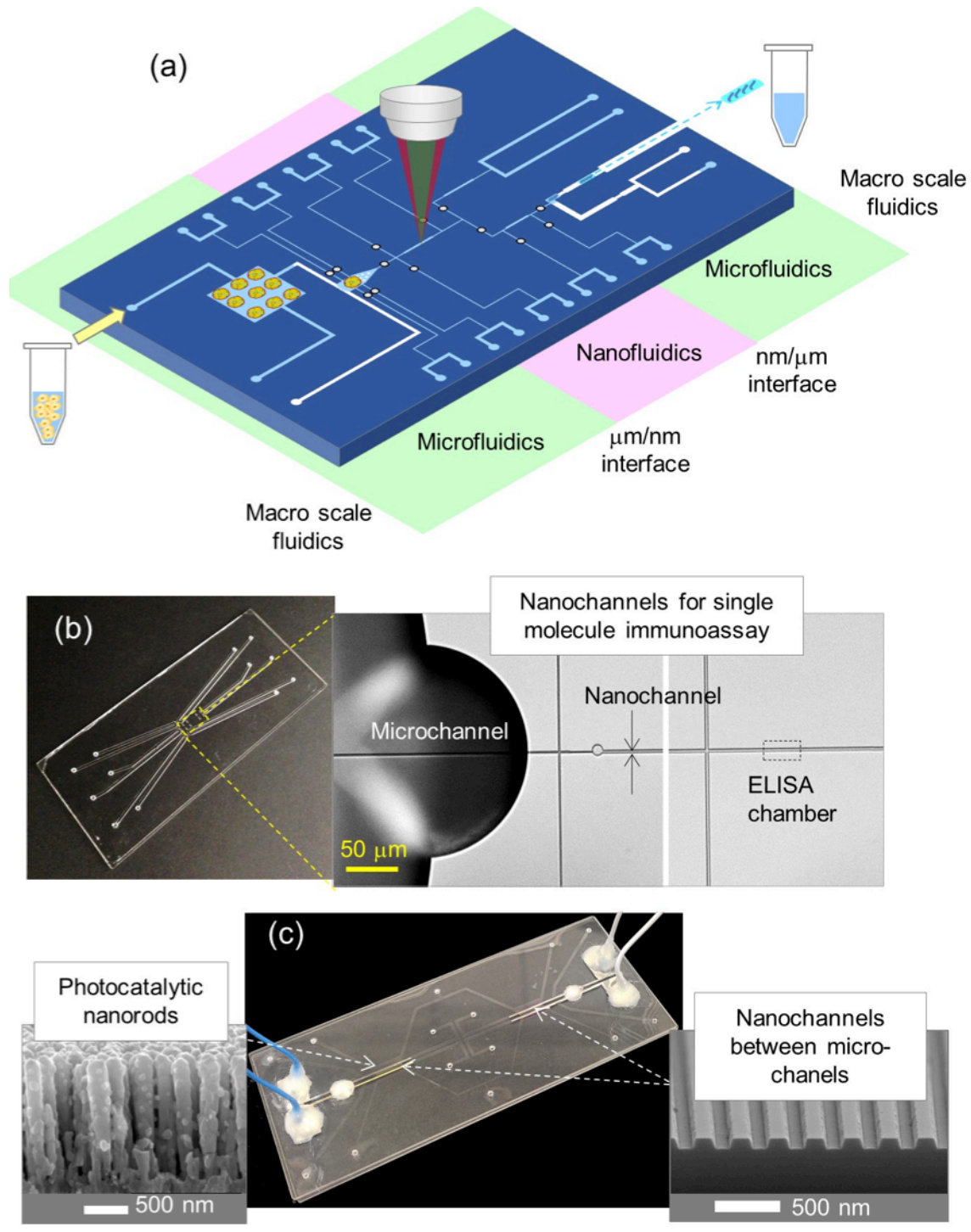

*Fig. 6. (a) Schematic of micro/nano hybrid fluidics. (b) Nanochannels connecting to the microchannel, which is for single molecule immunoassay at 100 femto liter sample. (c) Solar light driven self rechargeable m-fuel cell device by micro/nano hybrid fluidics. Left hand side is the fuel generator, and the right hand side is the fuel cell.*

It is difficult to access to nanochannels from our macroscale world, and a stepwise fluidic system from macro to nano via micro fluidics as shown in Fig. 6a is indispensable. This illustration is our strategy and platform of nanofluidics, and actually we realized these technological platforms and examples were demonstrated in Figs. 6b and 6c.

## 4. Breakthrough given by nanofluidics: A novel tool for science and technology

The ultimate smallness of nanofluidics enables chemistry and biomedical science at pico, femto, and even atto liter scale. Number of the target molecules becomes countable in this smallness. For example, we successfully demonstrated atto-liter chromatography separation, femto-liter and single-to-countable molecules immunoassay (Fig. 6b), and femto-liter solvent extraction. These applications are also combinations of nano unit operations, which is the same as MOU in microfluidics. Especially, in case of the immunoassay, the significant signals from single to countable antigen proteins were obtained and we proved the principle of single molecule immunoassay.

Unique properties and characteristics of liquid and fluid were found. For example, viscosity of water is several times larger than normal water, dielectric constant becomes 1/7, proton conductivity is two hundred times larger, fluid at the channel wall is slipping and its velocity is not zero, and so on as shown in Table 2. Water molecules in a glass nanochannel strongly interact with the channel wall, and the nature of molecule as a particle becomes remarkable in nano scale condensed phase.

*Table 2. Property changes of water in nanochannels.*

| Property | Change | Material | Size | Method |
|---|---|---|---|---|
| Viscosity of water | 1.4-4× | $SiO_2$, LBL | 25-1000 nm | Capillary force[1, 2, 3, 4] |
| Conductivity of water | 200× | $SiO_2$ | 100-1000 nm | Streaming potential[5] |
| Dielectric constant of water | 1/7-1/4 | $SiO_2$ | 330-800 nm | Fluorescence[1]<br>Streaming potential[6] |
| Proton mobility | 20× | $SiO_2$ | 40-800 nm | NMR[7] |
| Proton diffusion constant | 8× | $SiO_2$, LBL | 180-330 nm | Fluorescence[8] |
| Proton concentration (pH) | 19× | $SiO_2$, LBL | 400-1000 nm | STED[9] |
| Enzymatic reaction | 6× | $SiO_2$, LBL | 340-1000 nm | Fluorescence[10] |
| Dissociation of silanol | Accelerated | $SiO_2$ | 580 nm | Streaming current[11] |
| Vapor pressure of water | Decreased | $SiO_2$ | 120-510 nm | Microscopy[12] |

LBL: lipid bilayer

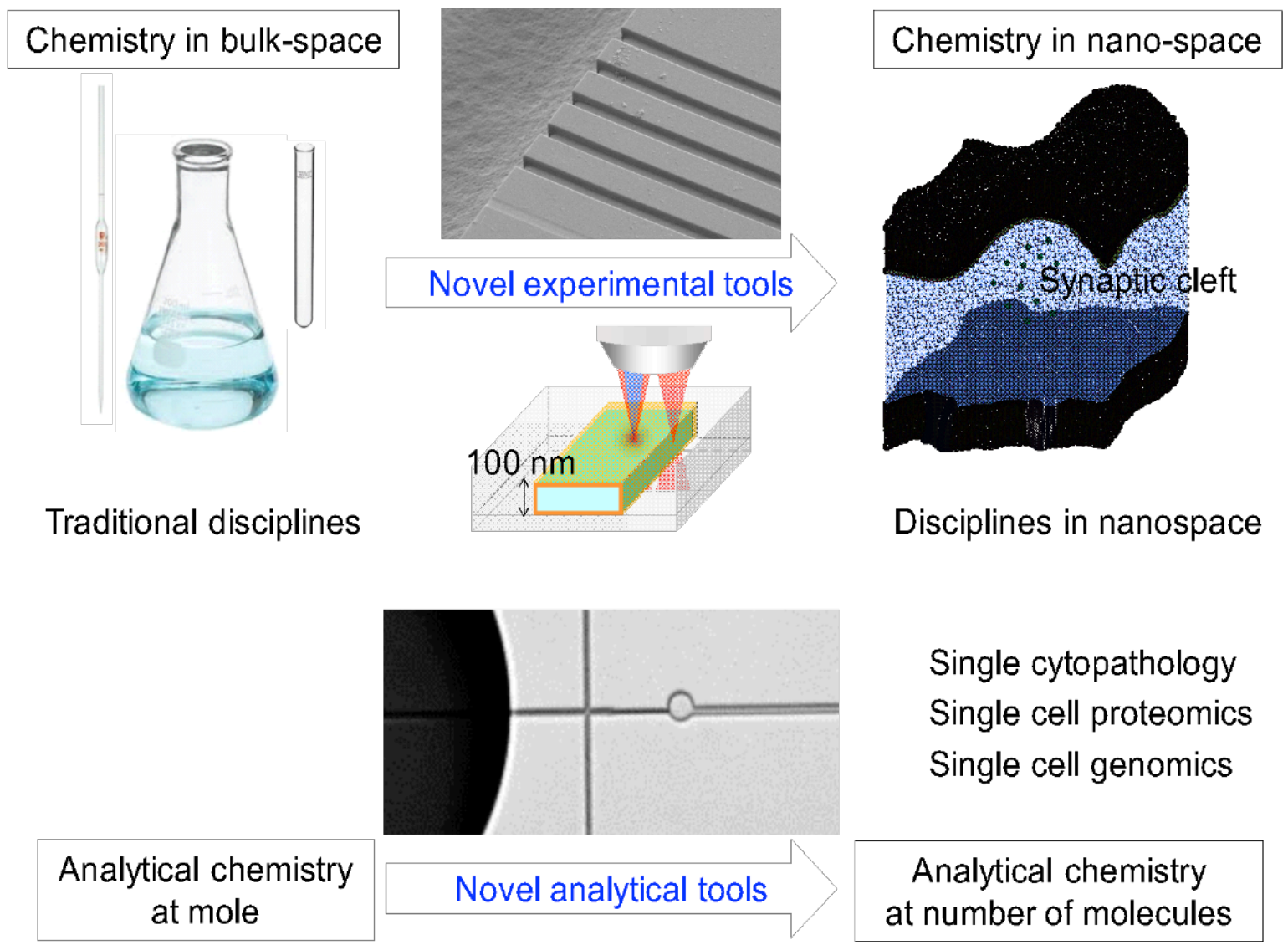

*Fig. 7. Nanofluidics in near future.*

In this way, nanofluidics is surely making progresses. We call our nanofluidics as "extended-nano fluidics" and will continue it, because we are interested in ultimatum of condensed phase and uninterested in isolated single molecules at single-digit nano space. Our intention is going well, and researchers having the same ambition and expectation are now making a community and sharing common prospects described in the next section.

## 5. Prospects

It is a matter of time that microfluidics brings breakthrough in many science and technology fields. Actually, the customers of our spin off company, Institute of Microchemical Technology (IMT), are now mainly companies and researchers of application side, although the main customers were microfluidics researchers ten years ago. But it has not made a big market yet. For the real implementation of microfluidics in industry and science, there are still some technologies to be developed. System technology, design method, and online flow control mechanism are the typical issues. However, we are feeling anticipation that microfluidics will break into a big market and contribute to science, industry, and society.

Regarding nanofluidics, the technologies at ultimate small volume and unique characteristics appearing in there may innovate chemistry from mole to number of molecules, and bring breakthrough tools to many kinds of science and technologies like biology in and between cells, pathology at cell level, fluid mechanics at molecular picture, and so on (Fig. 7). Actually, the young researchers in these fields are getting together in our lab, expecting to get a novel and powerful research tools.

Microfluidics and nanofluidics are worthy of paying attention to make sure that they will innovate science and technology and create new values in industry and society.

# Probing protein self-assembly and misassembly in small volumes

*Tuomas Knowles - University of Cambridge*

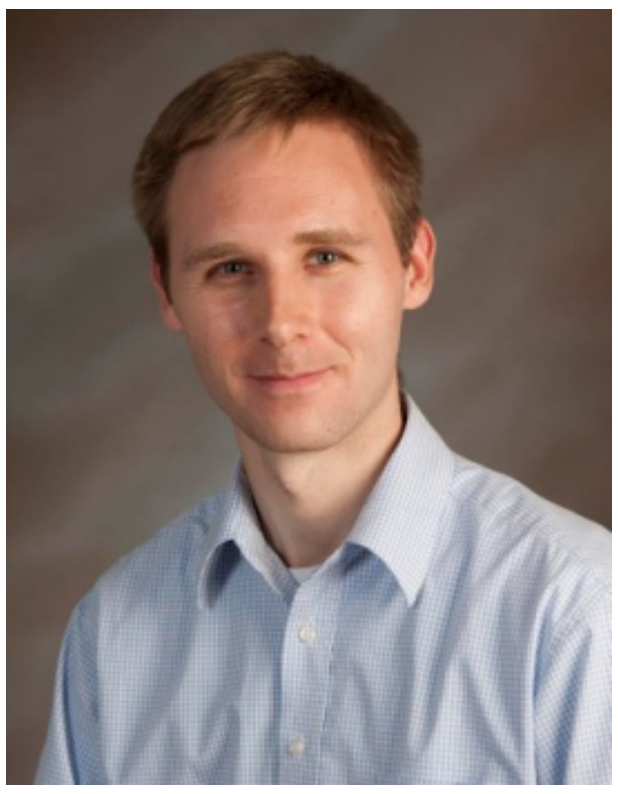

## Biography

*Tuomas Knowles studied Biology at the University of Geneva, and Physics at ETH Zurich, from where he graduated in 2004. He moved to Cambridge in 2004 as a PhD student working in the Cavendish Laboratory and the Nanoscience Centre. In 2008 he was elected to a Research Fellowship at St John's College, Cambridge, and was then appointed to a University Lectureship in Physical Chemistry in 2010, joining the faculty at the Department of Chemistry in Cambridge. He then successively held a University Readership between 2013 and 2015 and a Professorship since 2015 in the Department of Chemistry. Since 2016 he is Professor of Physical Chemistry and Biophysics in the Department of Chemistry and at the Cavendish Laboratory in the Department of Physics in Cambridge, and is co-director of the Cambridge Centre for Protein Misfolding Diseases. His research focuses on bringing together the development of new micro and nanoscale approaches to probe protein behaviour with key biomedical problems, including the molecular basis of neurodegeneration.*

## Protein self-assembly and misassembly

Protein molecules provide the functional basis for biological pathways in living systems. While a subset proteins are active as monomeric entities, for instance as enzymes, in the majority of cases, proteins are only able to fulfil their key biological roles in nature through self-assembly together with other protein molecules into supra-molecular structures that form the machinery of life. In cases where proteins bind an incorrect partner, misassembled and pathological supra-molecular structures can result.

A general class of such misassembled protein structure is that of amyloid aggregates. These species were initially discovered in the context of pathology where their formation is a hallmark of a range neurodegenerative disorders. These materials, including nanoscale amyloid fibrils, are characterised by the presence of a densely packed beta sheet rich hydrogen bonding network at their core, which confers a high level of chemical and mechanical stability and renders the assembly of these structures essentially irreversible under normal biological conditions, leading to the disruption of normal cellular pathways when their formation occurs in an uncontrolled manner in the context of disease. Moreover, lower molecular weight oligomeric amyloid species, formed from precursor peptides and proteins which can be entirely innocuous and normally present in the cytoplasm of cells, are in general potent toxins, and contribute significantly to the loss of cell viability in the context of neurodegnerative disorders.

Despite these key roles of supra-molecular protein self-assembly in both biological function and malfunction, this process has remained challenging to characterise in a quantitative manner. This situation arises partly due to the highly dynamic and heterogeneous nature of the species which are formed as a result of protein self-assembly, characteristics which render these systems less suited for study using well established tools of structural biology. Over the past ten years, we have focused our efforts into developing and applying new micro and nanoscale methods leveraging physics and engineering concepts in the context of protein science. This talk outline some results in the context of using micron scale flow control to probe protein self-assembly in the condensed phase in the natural environment of proteins, but at resolutions increasingly approaching state of the art gas phase mobility measurements. Moreover, we will discuss recent efforts that we have undertaken to extend the space of protein material architectures by bringing together protein self-assembly with microfluidic control of fluid flows.

## Probing protein self-assembly on small scales

We have focused our efforts on exploring experimental strategies to provide a new window into protein self-assembly that are enabled by operation in small volumes and that do not have a bulk equivalent. We have shown that microconfinement achieved through droplet microfluidics allows the isolation of single nucleation events in protein aggregation and thus to study a rare event as single molecule resolution. Using this strategy we have also been able to develop an understanding of how aberrant misfolded protein states are transmitted from one molecule to another through time and space. More recently we have exploited measurements of mass transport through fluid streams under laminar flow conditions to generate a platform for probing protein-protein interactions under fully native conditions.

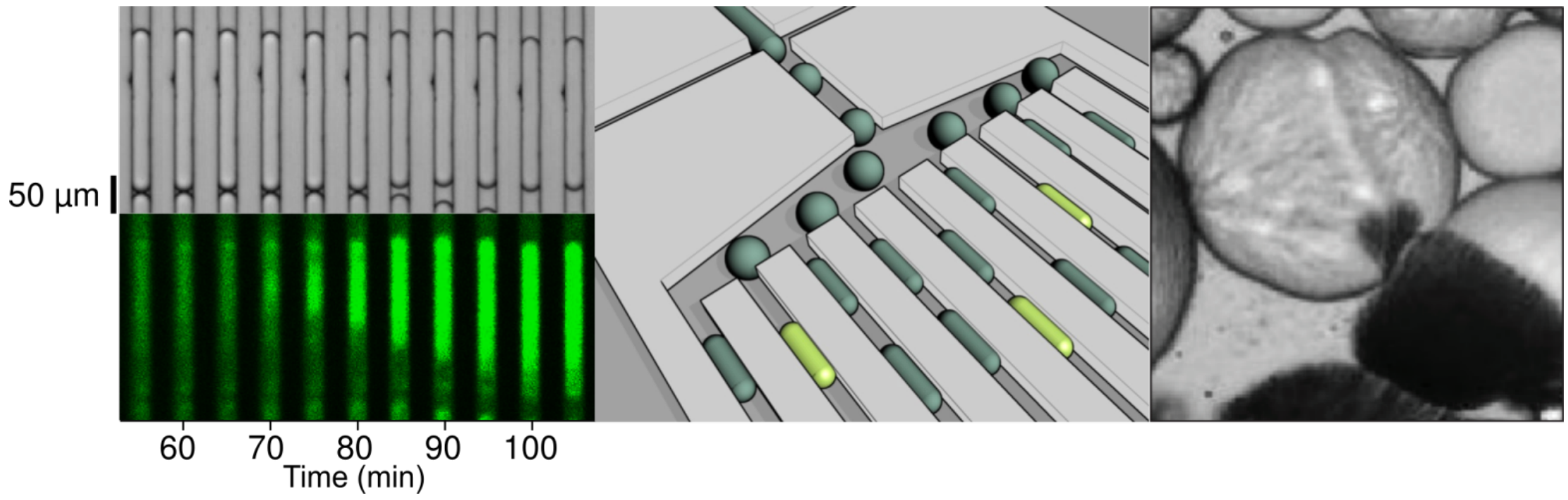

Figure 1: Use of confinement to probe protein self-assembly from single nucleation events (left and middle - PNAS (2011) 108, 14746) to force generation from self-assembly (right - Nature Physics (2016) 12, 926).

## Expanding the materials space accessible using protein structures

In nature, high performance materials, including spider silk and the cytoskeleton are formed from protein molecules through self-assembly. Capitalising on this natural ability of protein molecules to self-assemble to generate new artificial biomaterials from natural building blocks has been frustrated by the difficulty in controlling protein self-assembly to generate structures and materials outside of those selected by evolution. We have sought to address these limitations by bringing together filamentous amyloid protein assembly with micron scale control of fluid flows.

Amyloid fibrils are a common form of protein nanostructure resulting from the aggregation of soluble proteins into supramolecular polymers, which possess remarkable physical properties, including a high Young's modulus and tensile strength and the ability to self-assemble under mild conditions in aqueous solution. Such structures were initially identified in nature as pathological protein deposits, yet recently have emerged as key functional components in biological materials found in organisms ranging from bacteria to humans. However, the applications of the general nanofibrillar amyloid scaffold have been limited by the challenges in enabling the overall micronscale morphology to progress beyond spatially uniform gels. By combining the inherent nanoscale self-assembly process with micron scale structuring, afforded by droplet microfluidics, we establish a class of physical microgels based on nanofibril-forming proteins that are easy to synthesize, biodegradable and nontoxic and that show advantageous characteristics as vehicles for the encapsulation and release of active proteins such as antibodies or small molecules, including drugs.

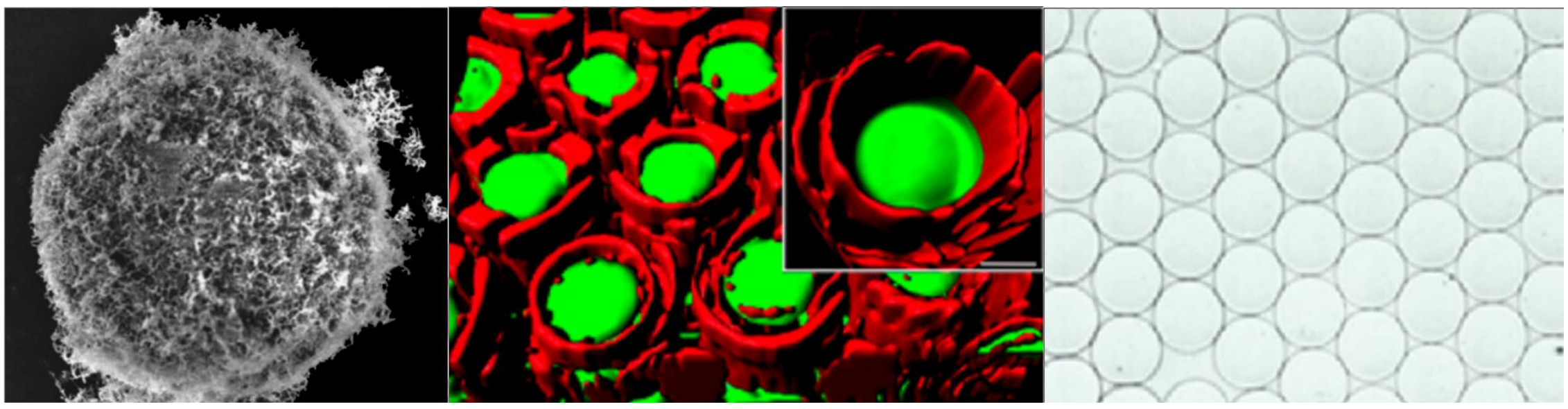

Figure 2: Artificial biomaterials from amyloid fibrils. We have approached the challenge of synthesising new functional materials from natural building blocks by bringing together protein self-assembly with microfluidic techniques to form protein microgels, shells and capsules, opening up novel material architectures not found for proteins in nature (Nature Communications (2016) 7, 12934, Nature Communications (2016) 7, 13190), Science (2007) 318, 1900, Science (2009) 326, 1533.

# Shrinking Genetic Analysis into a Portable and Cost-effective Microfluidic Form Factor

*James Landers - University of Virginia*

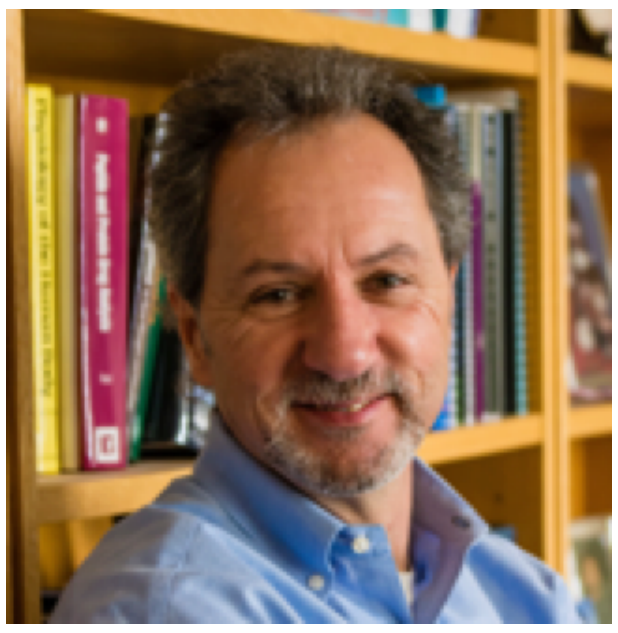

## Biography

*James Landers is a Jefferson Scholars Fellow and Commonwealth Endowed Professor of Chemistry, a Professor of Mechanical Engineering, and an Associate Professor of Pathology at the University of Virginia. James received a Bachelor of Science in Biochemistry (minor in Biomedicine) and Ph.D. in Biochemistry from the University of Guelph, Canada. He did a post-doctoral fellowship at the University of Toronto School of Medicine, and was a Canadian Medical Research Council Fellow in cancer biology at the Mayo Clinic where, In collaboration with Beckman Instruments, he launched and directed the Clinical Capillary Electrophoresis Facility in the Department of Laboratory Medicine and Pathology developing capillary electrophoresis-based clinical assays.*

*He joined the Division of Analytical Chemistry at the University of Pittsburgh in 1997, and began the development of analytical microfluidic systems for next generation molecular diagnostics; this continued at the University in the Virginia in the early 2000's. These efforts culminated in a 2006 Proceedings of the National Academy of Sciences report on one of the first sample-to-answer microdevices for genetic analysis; these early developments were incorporated into the Rapid DNA Analysis microfluidic system developed by MicroLab Diagnostics, a start-up biotech company. These microfluidic tools have been extended to a new DNA analysis platform that exploits centrifugal force to control fluid flow, chemical reactions and analytical processes – development of the faSTR DNA System is currently ongoing.*

*Landers group has published more than 240 papers in peer-reviewed journals, 25 book chapters and edited three editions of the Handbook of Capillary Electrophoresis, and his efforts in microfluidics led to the 2008 Association for Lab Automation 'Innovative Technology of the Year' Award. James Landers is currently the Microfluidics Editor for Analytica Chimica Acta, a Visiting Professor at the Kaohsiung Medical University (Taiwan), and the Director of the Center for Nano-Biosystems Integration at UVA.*

## Background

While next generation sequencing evolves rapidly, and will eventually become cost-effective enough to meet the bandwidth of genetic analysis needs, electrophoresis with multicolor detection still provides the foundational technology for many scientific arenas where comprehensive sequence information is not necessarily needed.

One application is DNA-based human identification (hID), where the length of a collection of tetra- and penta-nucleotide short tandem repeats (STRs) from different regions in the genome are unique to an individual; at the time of selection, these had no clinical relevance. The sizing of short tandem repeat sequences (STRs) from 13-24 genetic loci can, collectively, provide discriminatory power that allows the random match probability (RMP) that two individual have exactly the same STR profile to be on the order of one in trillions (Fig. 1).

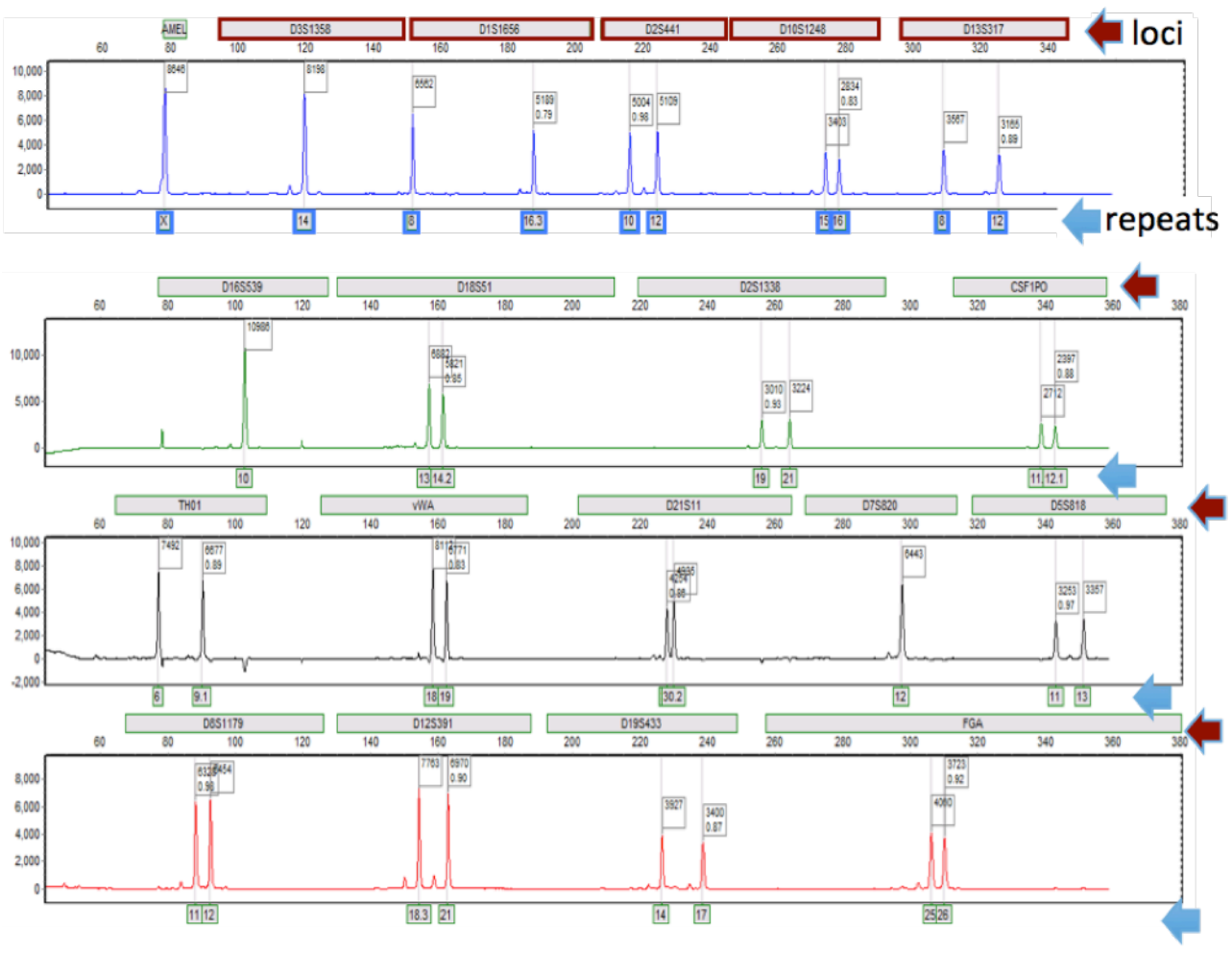

*Figure 1. A typical 18-locus profile used as court-defensible data with a RMP of one in 18 trillion.*

## 'Rapid' DNA systems

In 2008, the federal government funded the development of 'Rapid DNA' systems – genetic analysis systems that could provide sample-to-genetic profile in an automated format and in less than 90 min [J Forensic Sci. 58(4):866-74, 2013]. Several university start-ups were involved in developing commercializable systems specifically for generating STR profiles from buccal swabs, with the ultimate goal being to employ these systems in booking stations (with or without association with a forensic lab) throughout the USA . The nationwide DNA typing of convicted felons has been in effect for some time, but now 'Katie's Law, the routine collection of buccal swabs upon arrest, is in effect in more than half the states.

The 'Rapid DNA' movement blazed a new trail, an application-driven need that led to the integration of sample preparation (DNA extraction) and DNA amplification (PCR) steps with electrophoretic separation in a single microfluidic device – a bona fide lab-on-a-chip – for automated DNA processing (Fig. 2). It guided the criminal justice community into a new mindset, and forced the federal government to anticipate the policy changes that would be needed to allow booking stations to upload the generated STR profiles to the national database. However, to-date only a handful of booking stations have adopted the technology, and for substantive reasons. The capital investment is sizable at $250,000-$280,000 USD, exacerbated by a microdevice cost that, at $150-$500 per sample, makes

A

B

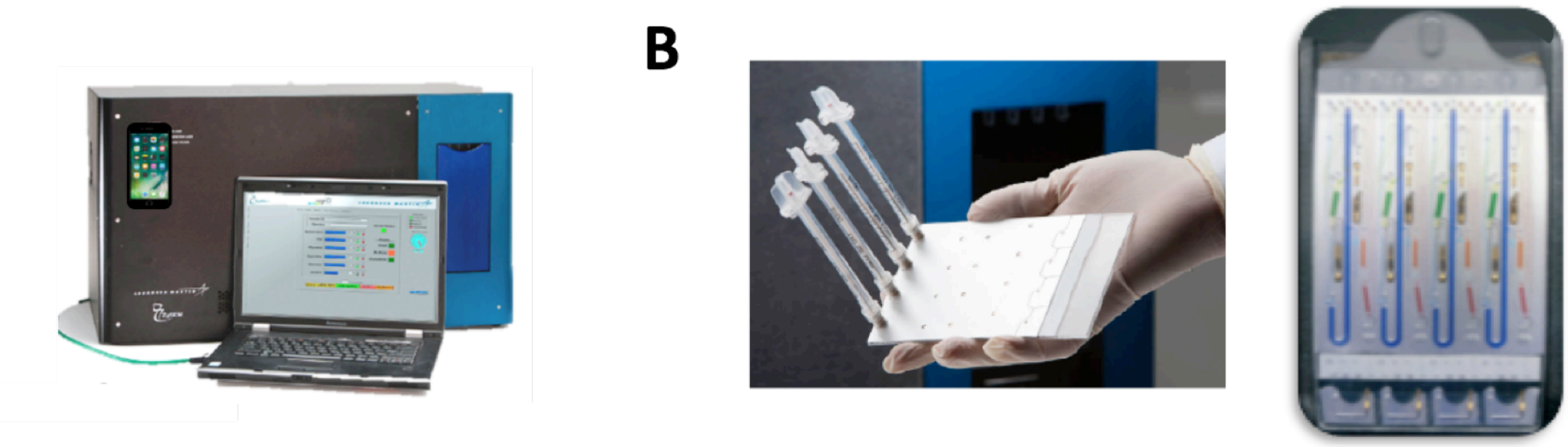

*Figure 2. 'Rapid DNA' system from one start-up company. (A) The IntrepID from MicroLab-ZyGEM (Lab Chip 14:4415–4425, 2014.) and (B) the MicroLab-ZyGEM chip capable of running 4 samples with swab-to-profile in 60 min. This chip has a completely self-contained fluidic architecture, can accommodate simultaneous analysis of 4 samples, involves a 5-layer assembly, has 2 independent fluidic layers accessible through a 'via' layer.*

‘Rapid testing’ almost an order of magnitude more costly than standard forensic lab analysis. Moreover, the size of the microdevice and/or the instrument size/weight has limited adoption of the ‘Rapid DNA’ technology by other sectors where lightweight, portable and cost-effectiveness were important factors.

## The faSTR DNA system

The faSTR disc is a 10-layer hybrid system composed of 7 different materials (COC, PE, PMMA, PSA, HSA, toner, gold leaf) with 24 chambers and 55 channels on 7 fluidic layers, with the fluid flow between layers and throughout the entire fluidic architecture driven by centrifugal force and governed by 13 physical valves needed to maintain fluidic isolation of the diverse chemistries associated with DNA extraction, PCR and electrophoresis (Fig 3-A & B).

The valves are opened by simple 2-sec irradiation with a 638 nm laser pointer (Fig 3Ci) controlled by the software and, guided by optical switch-controlled rotation with 0.1 degree accuracy and a radial accuracy of 20μm, hence, able to hit any 100 $\mu m^2$ area on the chip. All reagents are on-chip in lyophilized or blister pack form, including the polymer used for electrophoresis with 4-color fluorescence detection. Fluidic control is governed by software-controlled rotation speed/direction and the opening of valves. DNA extraction from cells collected on a custom ‘break-away’ swab is accomplished by mobilizing a thermophilic protease from the blister pack into the buccal swab chamber to generate PCR-ready DNA.

The sample DNA is mixed with on-board PCR reagents and thermocycling facilitated by ‘dual clamp-Peltier’. The target sequences for amplification include unique short tandem repeat (STR) chemistry for a gender marker (amelogenin) and 9 other loci defined as ‘core’ by the FBI. Following rapid PCR, the products are heat-snap cooled to assure the DNA denaturation single stranded form for electrophoresis. A novel non-crosslinked sieving polymer is mobilized from the blister pack into the electrophoretic domain using centrifugal force where, following sample injection, peerless resolution of DNA (2-base) in a short separation channel (5 cm) under modest electric fields (2KV) is completed in 300 seconds with fragment sizes out to 300 bases.

The instrument brings together a cadre of components required for disc spinning, heating, cooling, valving laser translation, clamping, optical detection and optics translation, along with the necessary custom-built software that controls all functions including post-electrophoresis deconvolution of the raw data to generate an ‘STR Profile’ (Fig 3D). Heating/cooling is accomplished with the dual Peltier system that facilitates extraction and PCR with temperature ramping at 10°C/second, while the detection optics interfaced with a blue diode laser that has an auto-alignment function, do not require a fluorescent-agent-filled channel. This last capability is unique because, unlike standard chip optics that are ‘hard-mounted’, a microdevice that is free-spinning must be optically aligned prior to electrophoresis to assure that detection is at the dead-center point of the 80μmx80μm electrophoresis channel. The entire system runs off of AC or battery, with the electrophoresis domain powered by a novel ‘step-up voltage’ system custom-built to generate 2000V from two 9V batteries.

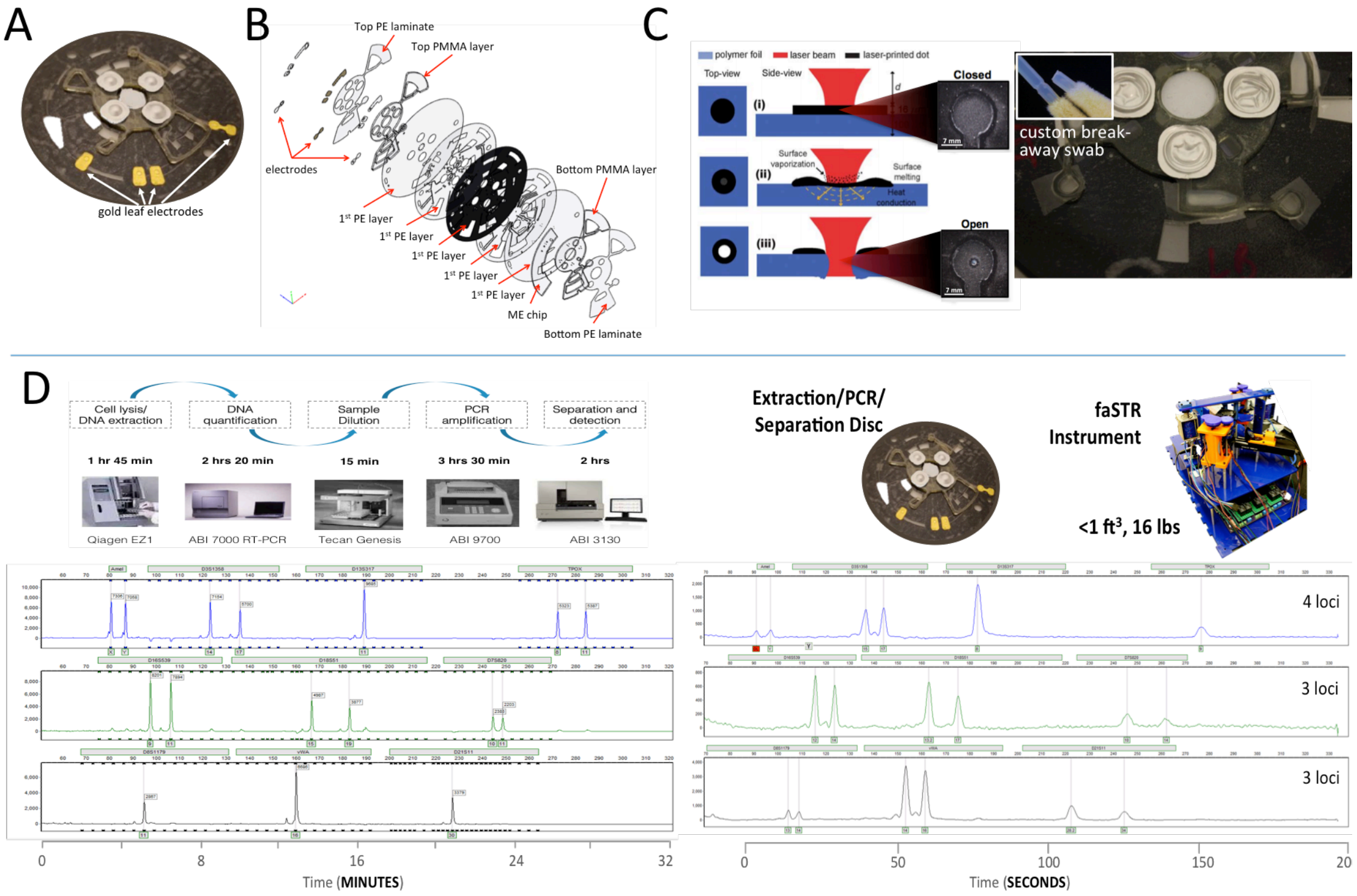

*Figure 3. (A) The faSTR microfluidic disc highlighting the gold leaf electrodes and blister packs. (B) Assembly of the integrated disc with the individual features needed with functionality. (C) Laser valving using an inexpensive red laser pointer and the custom swabs for sample insertion. (D) The faSTR instrument and comparison of a 10-locus STR profile generated by conventional methods (left) and the fully-integrated faSTR system (right).*

In a sample in-answer out data generated from a buccal swab in 60 min compares favorably with the data from conventional analysis (Fig 3D). The software guides the instrument through DNA extraction, PCR, electrophoresis and detection in an automated 27-step protocol with complete hands-off analysis and data generation in 60 min. This data is suitable for forensic human ID, but the application bandwidth is obvious, extrapolable to other points of analysis (POA). These include on-site analysis at mass disaster sites and for immigration control.

In summary, we have developed a portable genetic analysis system that is unprecedented in its capability, speed, size and weight, and eventually, cost. It breaks through the 'adoption barriers' set by 'Rapid DNA' technology and, perhaps, more important than all of the other parameters, is poised to bring a paradigm-shift to genetic analysis, as a result of its cost-effectiveness.

# Integrated sample handling in Biacore, a key component for the commercializing of micro fluidics based bioanalytical systems

*Stefan Löfås - GE Healthcare Bio-Sciences AB*

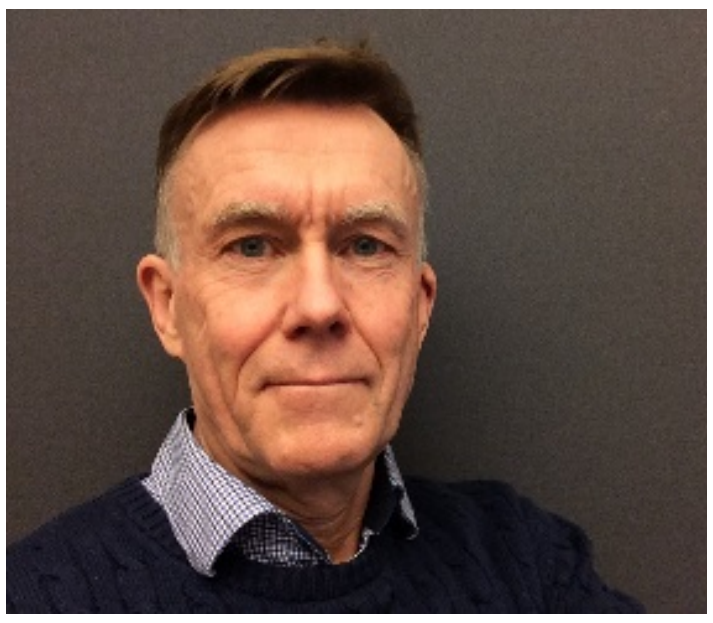

## Biography

*Stefan Löfås obtained his Ph.D. in Organic Chemistry in 1985 from Uppsala University in Sweden. After the dissertation, he took a position at Pharmacia Biosensor (in 1996 renamed to Biacore AB) as a research chemist and had a central role in the development of a new biosensor surface design and immobilization methods that where key elements in the launch of the Biacore™ system in 1990. He has since then hold different positions such as senior scientist and R&D manager in biochemistry and chemistry. In 2000 he was appointed CSO in Biacore International AB (acquired by GE Healthcare in 2006). He presently holds a position as Science Director at GE Healthcare Bio-Sciences AB with a focus on Strategic technology development within the Biotechnology area. Löfås is the author of several key publications on surface modifications and immobilization strategies for SPR based sensors and hold multiple patents within the SPR technology area. Löfås is also a member of the scientific advisory boards for the Chalmers Nano Technology Center and the advisory board of Center for Biomimetic Sensor Science at NTU, Singapore.*

## Abstract

One of the earliest successful innovations in advanced microfluidics relates to the development of the optical biosensor instrument Biacore. The company Pharmacia Biosensor AB (later renamed to Biacore AB and then acquired by GE Healthcare in 2006) in Sweden started research and technology development in 1985 and, thanks to major breakthroughs in surface chemistry, detector development and microfluidics based sample handling, could successfully launch the first in a long series of bioanalytical platforms in 1990.

The implementation of label-free, real-time interaction analysis based on the optical surface plasmon resonance (SPR) detection technology into a functional, user-friendly and quantitative analytical system was made possible by these technological developments. The SPR based Biacore systems has proven to be a versatile, general tool for studies of the affinity and kinetics of protein and biomolecular interactions within academic life science research, drug discovery and development, manufacturing and QC.

Such a technology intended to be used for quantitative interaction studies also put special demands on the precision and reproducibility of sample delivery in a miniaturized format to the sensor areas in the SPR detector. In addition, multiplexed detection by the utilization of several sensor areas will increase the information content and throughput, but have also to be matched to the needs for flexible addressing of the samples, in parallel or in sequence.

All these requirements emerged into the development of a novel type of miniaturized and integrated fluidics unit, controlled by pumps for liquid delivery over the sensor surface and with an autosampler handling module within the Biacore instrument for transfer from vials or microtitre plates (1).

The fluidics unit (or cartridge) is composed of micro-channels integrated with diaphragm valves that are pneumatically operated to facilitate rapid switching between buffer flow and delivery of samples as well-defined sample plugs (Figure 1). The valves are located close to the sensor areas, thereby reducing dispersion to a minimum. The flow channels are formed by precision casting in a hard silicon polymer plate. The thin-layer flow cells are formed by pressing the sensor chip against an area of the cartridge with multiple open rectangular grooves at the outlet side of the micro-channels. Typical dimensions for the flow cell are 1.6 mm long, 500 µm wide and 50 µm high, with a total detection cell volume of 40 nanolitres. The thin-layer flow cell enables optimal mass transfer conditions with linear flow rates from 1 to100 µl/min and under laminar flow conditions.

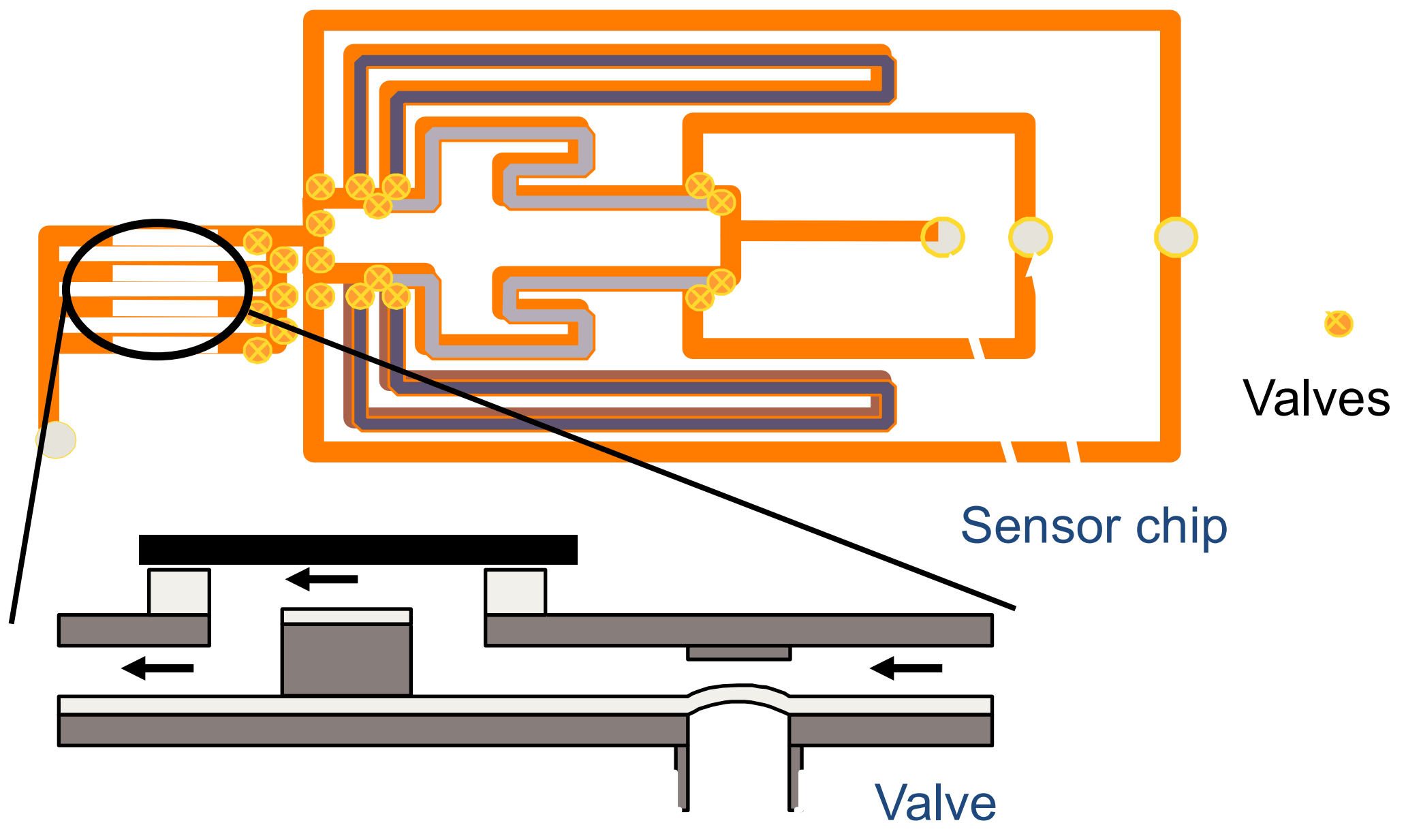

*Figure 1. Biacore microfluidic cartridge*

The fluidics cartridge is typically in operation in the instruments over a period of 6 months with up to 24/7 usage. The pneumatic valves can open and close hundreds of thousand times during that period at an air pressure of 250 kPa without any significant deterioration in performance. The material withstands all commonly used buffers and regeneration solutions for bioanalytical applications, including extreme pH conditions from pH 1 to 13.

Later versions of the fluidic cartridge have been developed for use in higher throughput versions of Biacore instruments. For example, a cartridge configured with a fluidic system with 4 parallel flow cells, each including five detection spots enables individual addressing of each spot by using hydrodynamic flow steering (Figure 2). This can be used for precise in situ immobilization of multiple ligands within different areas in the flow cell. This configuration also enables flexible assay configurations and facilitate generation of several hundred and up to thousands of data points in automated runs.

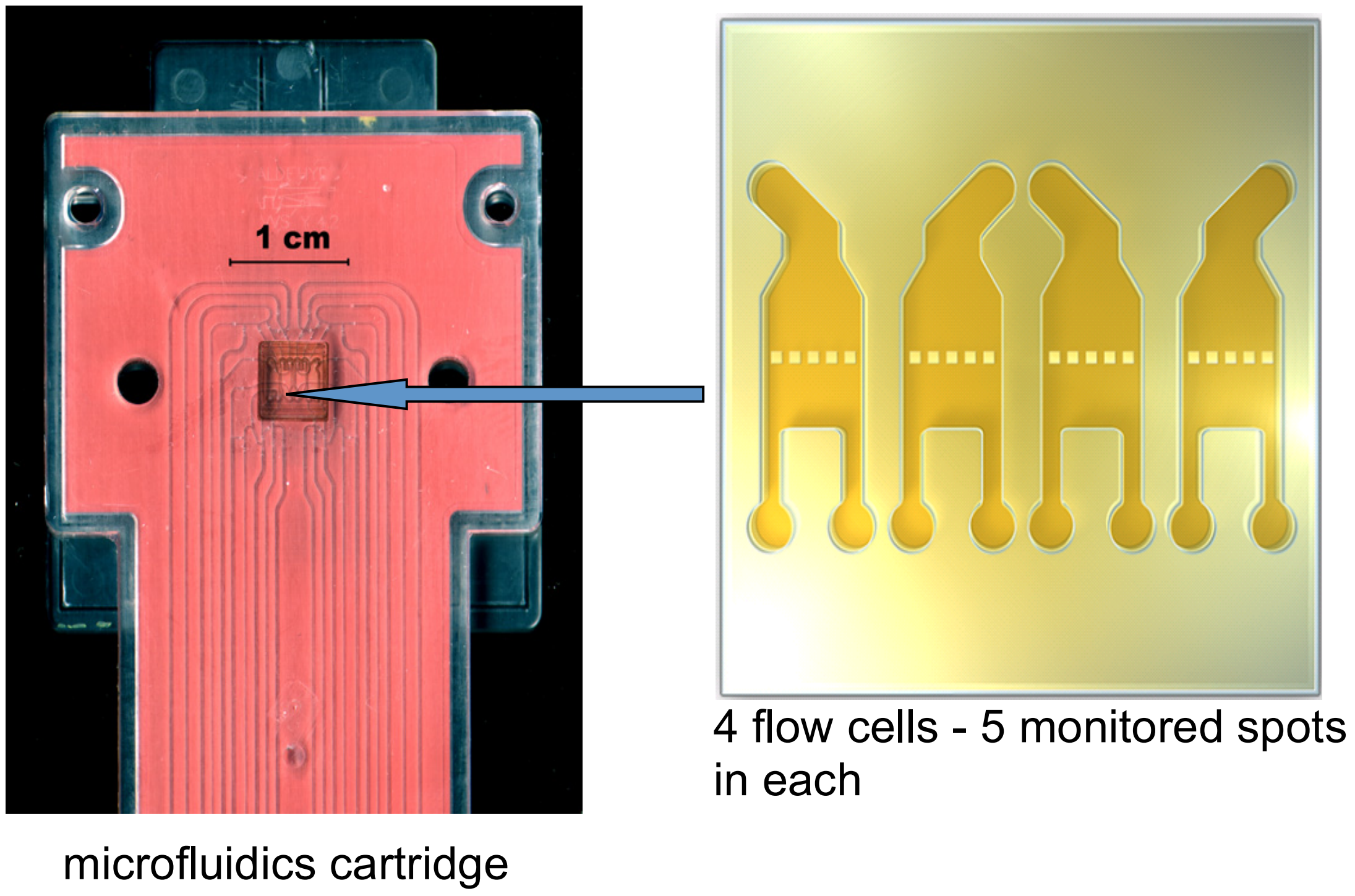

*Figure 2. Microfluidic cartridge for Biacore 4000*

There is no doubt that the main reason for the wide acceptance of the Biacore's technology has been driven by its use in quantitative kinetics and affinity analysis of biomolecular interactions within basic and applied research. This is evident from the steady stream of peer reviewed papers published since 1990 and with several thousand instruments on the market. Up to 2016 more than 15 000 papers based on data obtained on Biacore instruments can be tracked. It is evident that the implementation of the high quality integrated microfluidics has played a crucial role in the acceptance and commercial success of Biacore. This presentation will describe the historical aspects of this development as well as more recent advances of Biacore microfluidics and use in various applications.

# Microfluidics-enabled large-scale quantitative phenotyping

*Hang Lu - Georgia Institute of Technology*

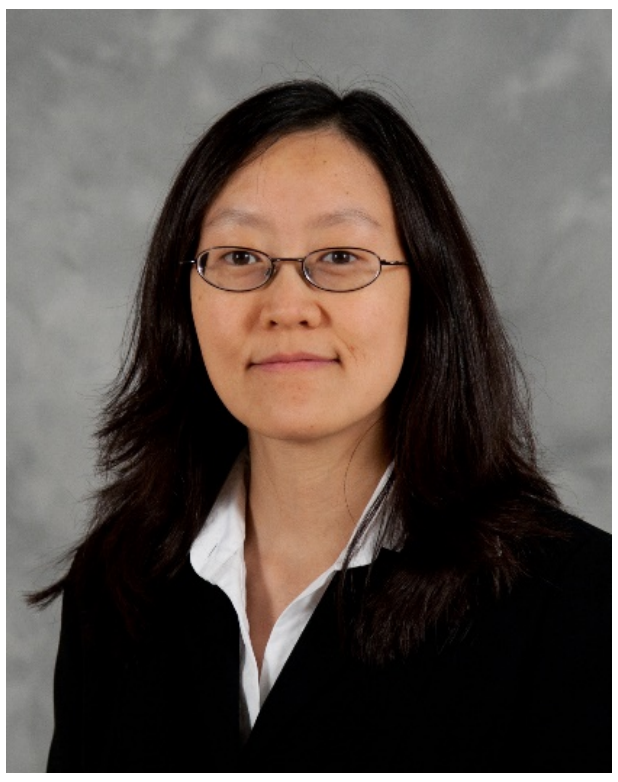

## Biography

*Hang Lu is the Love Family Professor in the School of Chemical and Biomolecular Engineering at Georgia Tech. She graduated summa cum laude from the University of Illinois at Urbana-Champaign in 1998 with a B.S. in Chemical Engineering. She has a Master's degree in Chemical Engineering Practice from MIT (2000). She obtained her Ph.D. in Chemical Engineering in 2003 from MIT working with Dr. Klavs F. Jensen (Chemical Engineering) and Dr. Martin A. Schmidt (Electrical Engineering and Computer Sciences) on microfabricated devices for cellular and subcellular analysis for the study of programmed cell death. Between 2003 and 2005, she pursued a postdoctoral fellowship with neurogeneticist Dr. Cornelia I. Bargmann (Howard Hughes Medical Institute investigator, Kavli Prize in Neuroscience 2012) at University of California San Francisco and later at the Rockefeller University on the neural basis of behavior in the nematode C. elegans. Her current research interests are microfluidics and its applications in neurobiology, cell biology, cancer, and biotechnology. Her awards and honors include the ACS Analytical Chemistry Young Innovator Award, a National Science Foundation CAREER award, an Alfred P. Sloan Foundation Research Fellowship, a DuPont Young Professor Award, a DARPA Young Faculty Award, and Council of Systems Biology in Boston (CSB2) Prize in Systems Biology; she was also named an MIT Technology Review TR35 top innovator, and invited to give the Rensselaer Polytechnic Institute Van Ness Award Lectures in 2011, and the Saville Lecture at Princeton in 2013. She is an elected fellow of American Association for the Advancement of Science (AAAS) and an elected fellow of the American Institute for Medical and Biological Engineering (AIMBE).*

## Summary of the talk

Microfluidics is a field of measurement science. Many successful microfluidic systems that have been developed in the last couple of decades have aimed at increasing throughput for measurements and enhance the precision of the measurements that are being made. In parallel, other microfluidic systems provide unique measurement opportunities and bring about new information otherwise unavailable through conventional assays. Development of new tools continue to change the way the measurements are made and used, to study a variety of biological questions. In this talk, I will focus on a few sets of tools my lab has developed in the context of organismal development, cellular function, and neural basis of behavior in small animals; I will also highlight the power of combining microfluidics with data science and automation.

## Cell trap arrays and other microfluidic devices for cell-based assays

Stimulating and imaging large number of cells over time, especially non-adherent cells are challenging. The Lu lab developed a cell trap array, which enables deterministic single-cell trapping and dynamic imaging of thousands of cells simultaneously while controlling their biochemical/physical microenvironment (Chung, Rivet et al. Analytical Chemistry 2011). This chip allows for thousands of cells to be imaged simultaneously, at single-cell resolution, while allowing for easy infusions of drugs or stimuli over the cells since microfluidics allows for simple integrations of complex upstream functionalities. Further variations of this device have also been developed (Chingozha et al. Analytical Chemistry 2014; He Kniss et al., Lab Chip, 2015; unpublished data), to allow complex stimulation patterns while performing live imaging, as well as end-point assays such as protein immunohistochemistry and in situ mRNA quantification, all at single-cell resolution. These techniques allow researchers to expand the type of phenotypical analyses that can be done on a single-cell level, for many cells, and correlate the phenotypes. We envision these systems to be of particular interest for cell biologists to study immunology, stem cell biology, and developmental biology.

## Microfluidics for studying embryos and development in vitro

Related to the cell traps, we have also developed a series of devices for larger samples such as Drosophila embryos and stem cell aggregates. The fruit fly Drosophila melanogaster is a model organism for embryo development and tissue morphogenesis. Typical studies of the dorsal-ventral patterning process require painful manual mounting and alignment of fragile embryos. The throughput of these experiments traditionally has been a few embryos per many months. The Lu lab has developed a microfluidic device for arraying, orienting, and high-resolution imaging of hundreds of Drosophila embryos (Chung, Kim et al. Nat. Meth. 2011), whose throughput is at least 100x faster than manual methods. This technology takes advantages of several hydrodynamic effects, together allowing an extremely high efficiency and success rate while avoiding complex instrumentation off-chip. Because of its facile utility, the chip has been used in many subsequent publications in collaboration with Shvartsman lab (Princeton U.) dealing with signal transduction in the ventral-dorsal patterning process (e.g. Kim et al. Developmental Cell 2011; Kanodia et al. Development 2011; Kanodia et al. Biophys J. 2012; Helman et al. Development 2012, Kanodia et al, Biophysical J. 2012; Lim et al PNAS 2013; Dsilva et al Development 2015; Samee et al Cell Systems 2015). Further development of this type of systems resulted in arrays to image embryos in different orientation and longer-term live imaging, as well as applications in cancer biology using Drosophila models (Levario et al Scientific Reports 2016; Goyal Levario et al, Disease Models & Mech 2017)

## Microfluidic systems for high-throughput screens and image-based genetics and genomics

The Lu lab works on micro technologies to revolutionize microscopy, phenotyping, and visual screens of multicellular organisms such as the nematode *C. elegans*. Imaging is extremely powerful in understanding how complex processes are orchestrated *in vivo*. Currently standard technologies widely used in biology laboratories and drug discovery processes suffer the drawbacks of being low-throughput, manual, and non-quantitative. The Lu lab designed a series of microfluidic devices image markers (e.g. fluorescent transcriptional markers or fusion proteins) for genetic screens, where the integrated system takes in a mutagenized population and is designed to have mutants sorted based on altered morphometric phenotypes. The lab has developed several strategies to immobilize worms, including cooling (for imaging marker that do not change rapidly, Chung et al, Nat. Meth. 2008), geometrical constraint (for non-stringent immobilization criterion, Crane et al. Lab Chip 2009, Caceres et al, PLoS ONE 2012), and using a biocomplatible gel that transitions from solution to gel phase upon 2 °C temperature change (for imaging at physiological conditions, e.g. imaging synaptic vesicle trafficking and calcium dynamics, Krajniak et al. Lab Chip 2010). These devices streamline the handling and examination of *C. elegans*, and allows for morphometric screens to be performed much more easily than manual methods.

Furthermore, the Lu lab has also focused on automation, image processing, and classification, which are integrated with the microfluidic chips to achieve the high-throughput experimentation. These techniques address one of the significant drawbacks of manual screening approach, which is the fact that phenotypes are identified subjectively by a user. This approach is sufficient in experiments where phenotypical changes are dramatic. However, most experiments using fluorescent markers now look for subtle changes, such as slight changes in intensity, location, distribution, size, and shape. By using computer vision, quantifications of these features on the fly are now possible and the classification of the sample much more accurate. For example, the Lu lab developed a set of machine vision techniques to identify synapses from background fluorescence, and use this information to find new mutants with altered synaptic marker localization (Crane et al. Nat. Meth., 2012). More recently, the Lu lab has developed yet another set of data-driven computational tools to understand genotypical-phenotypical relationships (San Miguel, Nat. Comm. 2016). Phenotypical heterogeneity in isogenic populations greatly reduces our ability to detect subtle mutants; however, by exploiting the power of the high-throughput microfluidic systems and statistical approaches, we show that we can resolve extremely subtle and complex differences between genotypes. Further, deep phenotyping using these complex traits allow us to gain an understanding of how genetic networks influence phenotypical manifestation.

## Microfluidics-based behavioral and neural imaging assays to study sensory neurobiology

Another area that the Lu lab is interested in is quantitative studies of neural or genetic basis of behavior. The technical challenge with behavioral studies is that most assays are qualitative. For instance, despite the heroic efforts in chemical identifications of various pheromone molecules, not much is known about how these molecules are sensed and how chemical communication is achieved using a combination of these small molecules. Behavior assays typically performed with these compounds usually require large amounts of chemicals and manual manipulations and tracking of animals. The Lu lab developed a microfluidic device to house many C. elegans in parallel but individually (Chung, Zhan et al. Lab Chip 2011). Chemicals can be delivered with temporal accuracy and behavior can be tracked and analyzed by image processing code automatically. The device uses only a fraction (~1/100) of the chemicals required to perform these assays and can be potentially used for long-term culture and assay of behaviors in worms.

Recently, the Lu lab has developed a microfluidic platform for delivering robust and precise mechanical stimuli to *C. elegans* by using pneumatically actuated structures (Cho Porto et al Lab Chip 2017; unpublished data). The device is fully automated, minimizing human variability and improving experimental throughput; it is fully compatible with fluorescent imaging of calcium dynamics of neurons. We show that this device can deliver mechanical stimuli with precise control in timing, location, and deformation. These characteristics allow us not only to recapitulate the well-known receptive field of the neurons in the gentle-touch circuit, but also to understand the dynamics of the harsh-touch neurons. We further developed this microfluidic system to interrogate behavior of mechanosensory neurons during development (both in active larvae and in larvae in lethargus state). In a collaborative effort, we are examining how small molecules may affect the signal transduction properties of human Transmembrane channel-like proteins in heterologous systems. Our microfluidic system directly addresses the current bottleneck of a lack of functional assay for such drug screens. The ability to automate and streamline the microfluidic system will greatly enhance our ability to identify potential therapeutics.

In summary, we show that microfluidics is instrumental in gathering biological information difficult or impossible to do with conventional techniques, and in studying genes and pathways in fundamental biological processes, which may be potential targets for future therapeutic development. We believe good designs are grounded in understanding of the fundamentals. Robustness and demonstrations of success are key to useful technologies for scientific discoveries. The future may lie in integration with other aspects of measurement sciences, such as quantitative analysis, particularly using statistical approaches.

This work is supported by funding from the US National Institutes of Health, the US National Science Foundation, and the Love Family Foundation.

# “Lab on chip” – the playground is over – where the playground now is

*Andreas Manz, - KIST Europe*

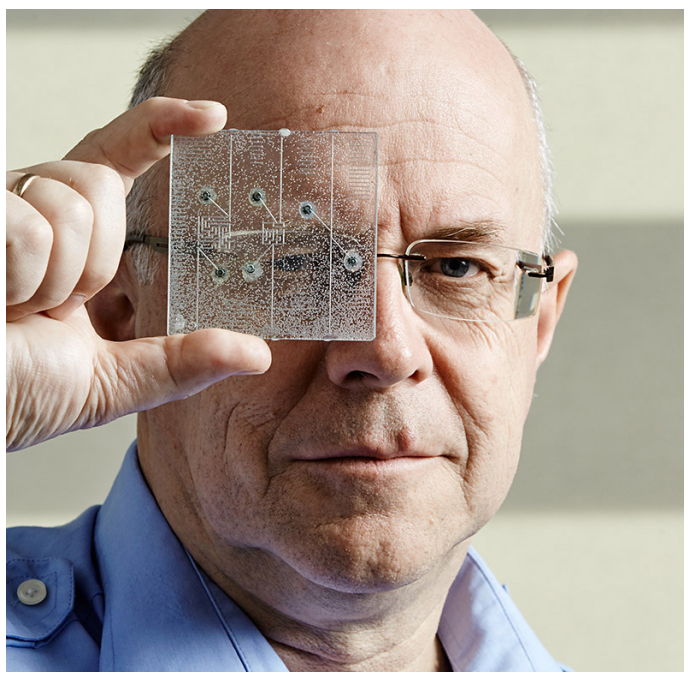

## Main research and management experience

- *Since 2010 professor of microfluidics for the life sciences, Mechatronics Department, Saarland University, Saarbrücken, Germany*
- *Since 2009 KIST Europe, Korea Institute of Science and Technology, Saarbrücken, Germany*
- *2009-10 visiting professor, FRIAS, fellow of the school of soft matter research, Freiburg Institute for Advanced Studies, Albert-Ludwigs-University Freiburg, Germany*
- *2003-08 head, ISAS - Institute for Analytical Sciences, Dortmund and Berlin, Germany*
- *1995-04 SmithKline Beecham professor of Analytical Chemistry at Imperial College, Dept. Chemistry, London, U.K.*
- *1988-95 Ciba-Geigy Corporate Analytical Research, Basel, Switzerland.*
- *1987-88 postdoctoral fellow at Hitachi Central Research Lab., Hitachi Ltd., Tokyo, Japan.*

## Education

- *1983-86 Dr. sc. tech. thesis (PhD) under the guidance of Prof. Dr. W. Simon (Dept. for Organic Chemistry, Swiss Federal Institute of Technology, ETH, Zürich, Switzerland): 'Small Volume Potentiometric Detector for Open-Tubular Column Liquid Chromatography'.*

## Honors and awards

- *2017 MNE Fellow Award, ecoplus, Business Agency of Lower Austria, Vienna*
- *2017 Martin Medal, Chromatographic Society, London UK*
- *2016 Chairman’s Award, National Research Council of Science & Technology, Korea*
- *2015 European Inventor Award, category ‘lifetime achievement’, European Patent Office, Paris*
- *2015 awarded the International Solvay Chair in Chemistry, The International Solvay Institutes for Physics and Chemistry, Bruxelles*
- *2015 Khwarizmi International Award, IROST, Iran*
- *2008 fellowship of the Koninklijke Nederlandse Akademie van Wetenschappen, The Netherlands*

## Abstract

The idea to shrink a laboratory down to chip size, or at least to integrate some critical chemistry lab tasks onto micro devices has been around since almost 30 years now. A 10x smaller scale means 100x faster throughput for chemical reactions or separations, if diffusion controlled. In the meantime, the field has grown significantly, many of my students are now professors in this field, the early patents are all expired, small companies use microfluidics for their products, and the main application focus has shifted from analytical chemistry to cell biology and tissue engineering. But where are the throughput benefits for chemical reactions or bioassays? [1]

I intend to show a personal perspective of all this, as I was involved since the early days. The case of electrophoresis on chip, from scaling laws to applications and commercialization will be shown. Drug discovery, combinatorial synthesis or clinical diagnostics could be applications, but just a fraction of it is actually in use.

Currently, the playground appears to have moved to different fields of interest. From my lab, I might show developments of remote controlling individual magnetotactic bacteria, of scaling self assembly concepts and of biomimetic structures obtained from nature.

Microfluidic chips are usually defined by photolithography masks which are generated from straight lines and CAD programs. The manufacturing process needs clean room technology and usually gets more complex if multiple depths, i.e. multiple masks have to be used, and variations in depth profile are difficult to achieve. I will present a simple way of obtaining channel structures which feature gradually increasing or decreasing channel depths, and which also can feature irregularities in its surface. At first sight, this may seem inappropriate, may look “ugly” and not engineering-like. However, in biological surroundings, we can see such structures, and they are fully functional.

Plant leaves are used as templates for channel patterns, including its fine structure and including its macroscopic network pattern. Structures are formed in PDMS and covered by glass slides for microscopic observation [2]. Structures are used for investigating cell behaviour, digital PCR and perfusion of 3d cell cultures.

# In vitro Brain Disease Models on a Chip

*Jianhua Qin - Chinese Academy of Sciences*

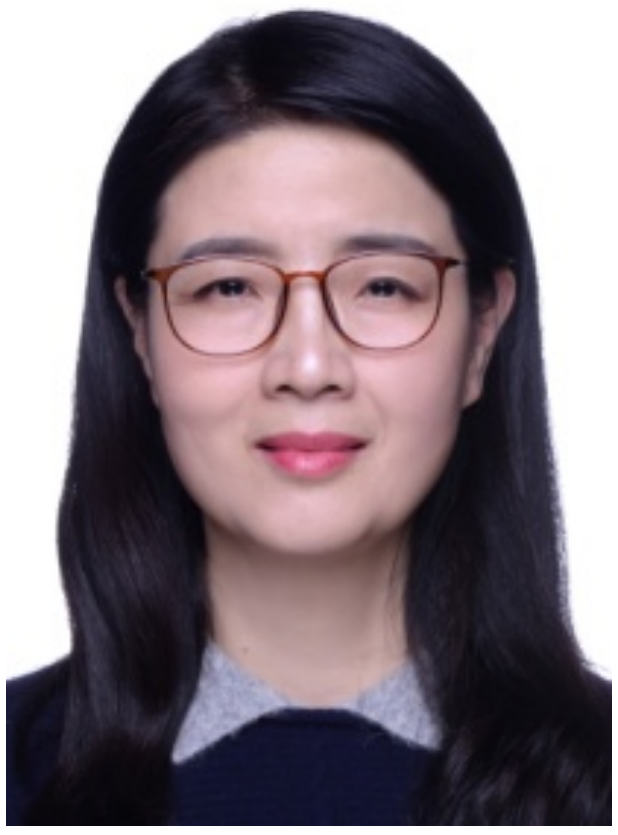

## Biography

*Dr. Jianhua Qin is a Professor at Dalian Institute of Chemical Physics (DICP), Chinese Academy of Sciences, and Director of Microfluidics Research Center in DICP. Dr. Qin received her Ph.D in Chemistry at CAS, Chinese Academy of Sciences and her M.D in Clinical Medicine from China Medical University. She was a visiting professor at the University of Hong Kong and the University of Toronto. Prior to this, she was an associate professor at the First Affiliated Hospital of Dalian Medical University, China. Dr. Qin has authored more than 100 peer-reviewed papers and issued 40 patents to date. Dr. Qin is the fellow of Royal Society of Chemistry and the Associate Editor of Lab on a Chip.*
*Her main research interest lies in the interface between microfluidics and biomedical sciences to understand human physio-pathology that leads to the design of new diagnostic schemes and therapeutic strategies.*

## Abstract

Brain diseases are among the most serious health problems facing our society, causing massive human suffering and economic costs. The lack of understanding the underlying disease mechanisms is a major obstacle to the development of better treatments. Pathological perturbations of the brain often involve the altered environments and the extraordinary complex, yet highly disorganized neural architecture in human brain. As such, establishing *in vitro* brain disease models is critically important to improve our understanding of the central nervous system, diseases pathology, potential neurotoxic effects of environmental factors, and the development of effective treatment. In this talk, I will present our efforts to establish several near-physiological brain disease models using microfluidic approaches combining with classical cell biology strategy. I will also demonstrate the utilities of microfluidic platforms for applications in probing the cellular pathology and mechanisms underlying brain tumor metastasis, and brain developmental disorders after various chemicals exposures.

## 1. Engineering the *in vitro* blood-brain-barrier microenvironment to probe brain metastasis

Evidence suggests that the physical, cellular, and non-cellular microenvironment of defined anatomical regions within the brain may contribute to the aggressiveness and resistance to the treatment of brain tumors. The blood brain barrier (BBB) is a dynamic interface between the blood and brain in vivo. The dysfunction of this barrier is closely associated with the brain tumor. The BBB restricts the uptake of many neuro-therapeutic molecules, presenting a formidable hurdle to elucidate the mechanism and drug development in brain diseases. We proposed a new and 3D dynamic microfluidic system that replicates the key structural, functional and mechanical properties of the BBB in vivo (Fig 1). Multiple factors in this system work synergistically to accentuate BBB-specific attributes-permitting the analysis of complex organ-level responses in both normal and pathological microenvironment in brain tumors. The complex BBB microenvironment is recreated in this system via physical cell-cell interaction, vascular mechanical cues and cell migration. This model possesses the unique capability to examine brain metastasis of human lung, breast and melanoma cells and their therapeutic responses to chemotherapy. The results suggest that the interactions between cancer cells and astrocytes in BBB microenvironment might affect the ability of malignant brain tumors to traverse between brain and vascular compartments, providing a versatile platform for drug testing and brain disease studies.

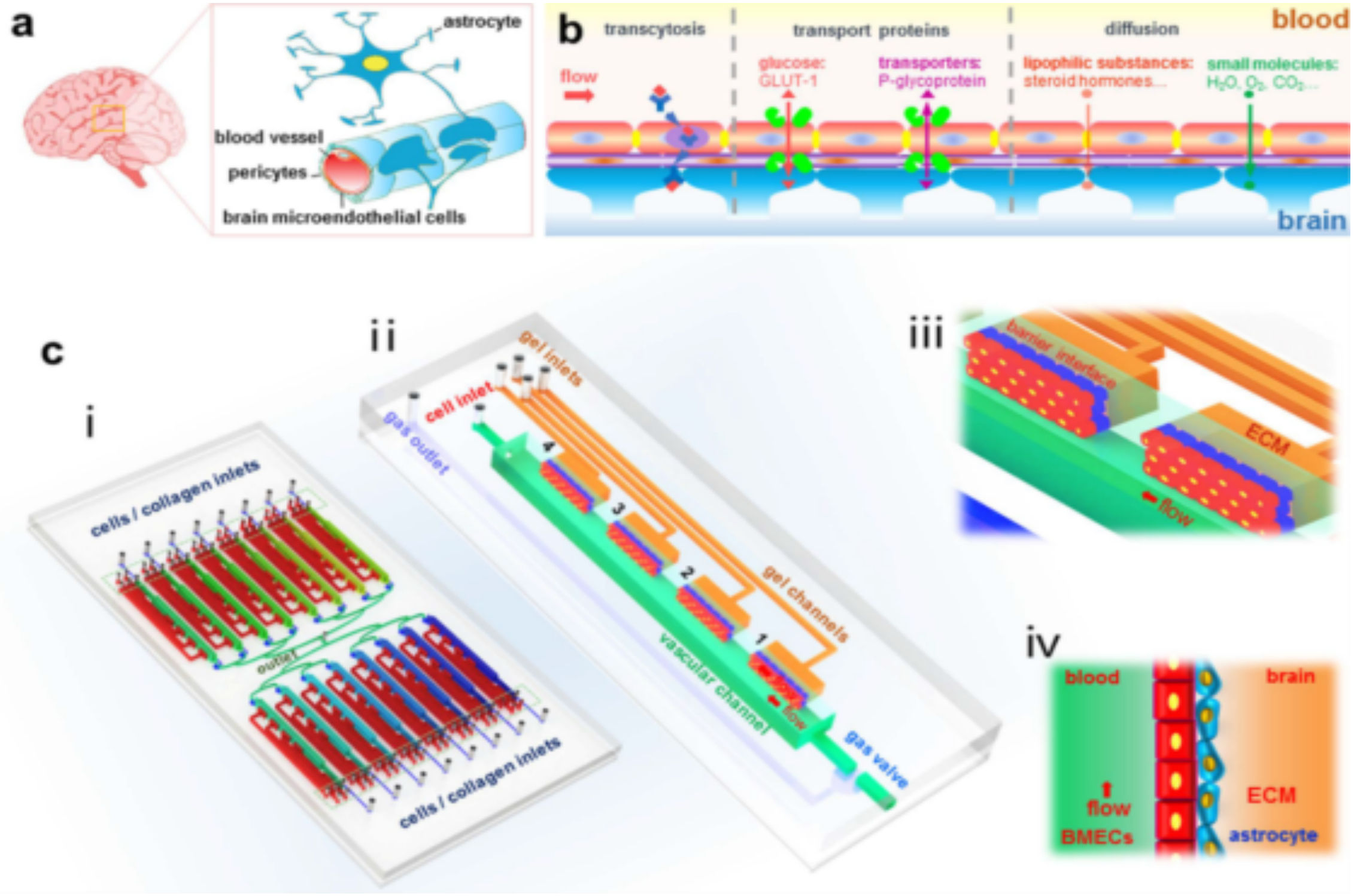

*Fig 1. BBB-on-a-chip*

## 2. Engineering stem cell-based brain organoids to study brain developmental disorders

The human brain is susceptible to many different disorders that appear at every stage of life. The fetal brain is highly vulnerable to various environmental exposures, which can trigger various long-term neuronal disabilities and cognitive dysfunctions. However, a comprehensive understanding of fetal brain development under toxic exposures is challenging due to the limitations of animal models. As such, we proposed a novel microfluidic strategy by combining with development biology to explore the mechanisms underlying neural dysfunctions with various environmental exposures, such as ethanol and nicotine. The functional microfluidic device enables the generation of human induced pluripotent stem cells (hiPSCs)-derived brain organoids in a simple and efficient manner (Fig 2). The 3D brain organoids recapitulate human brain organogenesis at early stages during gestation, including neuronal differentiation, brain regionalization, and cortical organization. With exposures, the brain organoids displayed skewed neurogenesis and persistent abnormity in neural maturation. A series of markedly altered genes and enriched pathways related to neurogenesis, craniofacial development and brain volume contribute to our understanding of the various postnatal neural disorders as observed in brain development disorders.

Overall, our work described the creation of in vitro brain disease models by incorporating the key essentials of brain microenvironment and brain organogenesis into the microfluidic devices. These work meld the concepts and engineered techniques with classic cell biology and developmental biology strategies to exploit the pathophysiology of brain diseases. The microfluidic technology is expected to provide a powerful tool to investigate cellular and molecular mechanisms underlying various brain disorders and to guide effective treatment.

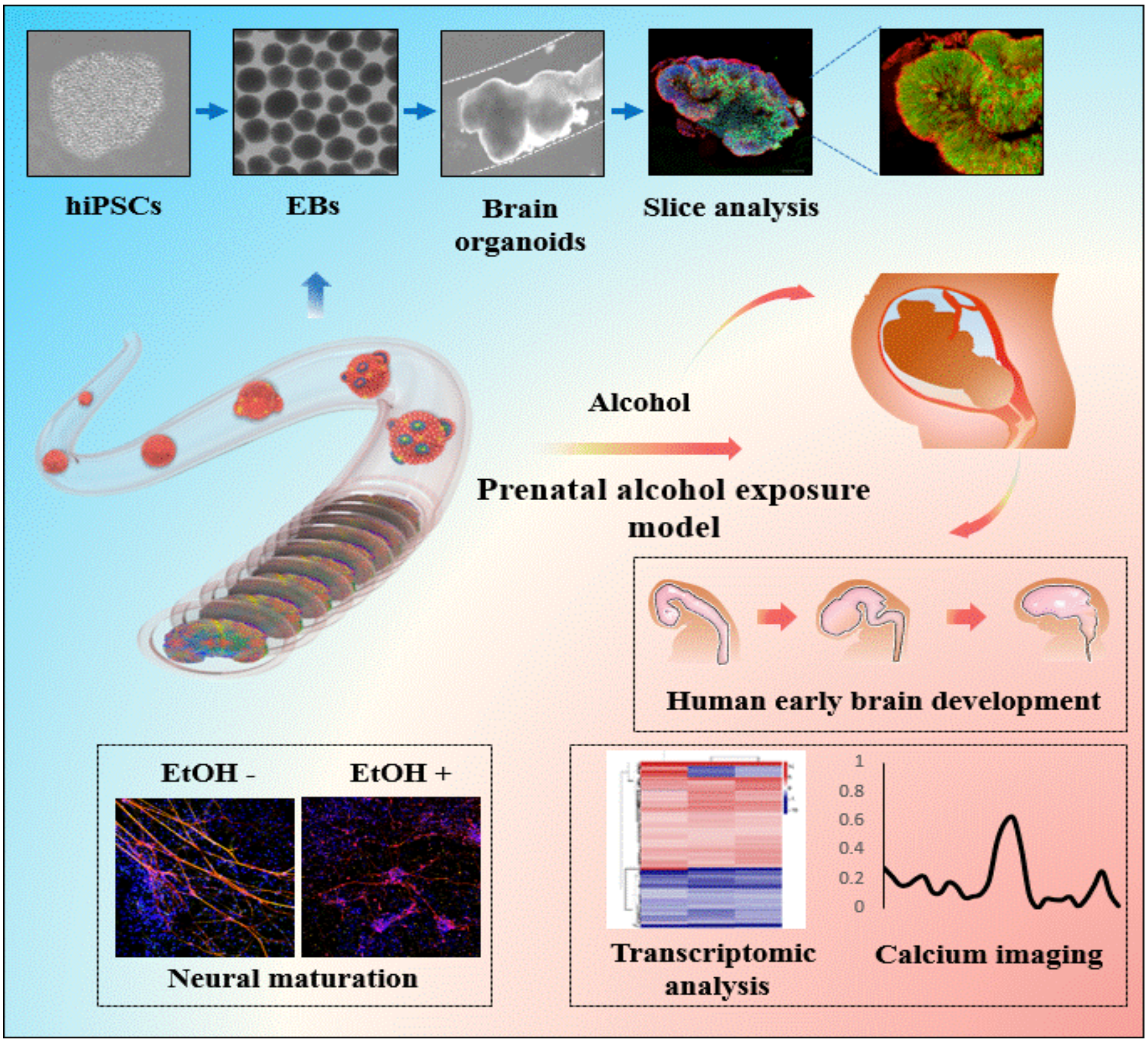

*Fig2. Brain organogenesis-on-a-chip*

# Drop-Based Microfluidics: Fundamentals and Applications

*David A. Weitz - Harvard University*

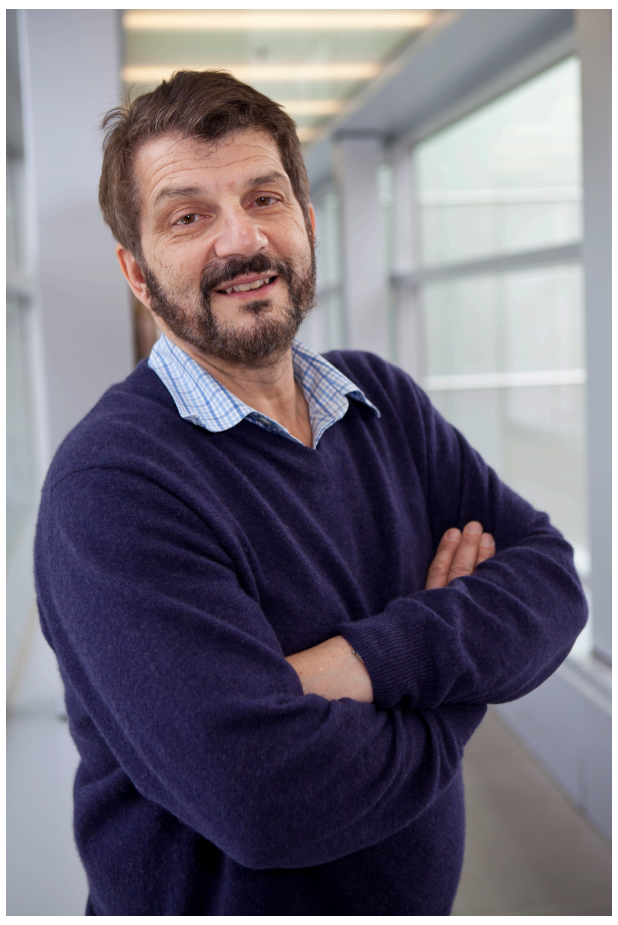

## Biography

*Weitz received his PhD in physics from Harvard University and then joined Exxon Research and Engineering Company, where he worked for nearly 18 years. He then became a professor of physics at the University of Pennsylvania and moved to Harvard at the end of the last millennium as professor of physics and applied physics. He leads a group studying soft matter science with a focus on materials science, biophysics and microfluidics. He is a core faculty member at the Wyss Institute for Biologically Inspired Engineering. He is director of Harvard's Materials Research Science and Engineering Center, funded by the National Science Foundation. Several startup companies have come from his lab to commercialize research concepts. Weitz is a member of the National Academy of Sciences, the National Academy of Engineering, and the American Academy of Arts and Sciences.*

## Abstract

This talk will provide a brief overview of drop-based microfluidics, with a discussion of the fundamentals of drop formation, and the use of the technology for both production of new materials and for 'lab-on-a-chip' applications.

The first use of drop-based microfluidics that I will describe is for production of new materials. Microfluidic devices enable precise control of the flow and mixing of fluids to produce drops of one fluid in a second, creating an emulsion. They can also produce drops within drops to create more complex, multiple-emulsion structures. This talk will review capillary microfluidics, a very simple means of creating microfluidic devices that can produce multiple emulsions. Exquisitely precise new structures can be created that possess both new properties and new utility. The talk will describe the production and properties of such multiple emulsion drops. In addition, the talk will describe methods to up-scale the production of these materials for practical materials applications.

The second use of drop-based microfluidics that I will describe is the 'lab-on-a-chip' applications for biology and biotechnology. Aqueous drops whose volume is between about one picoliter and about a nanoliter can be produced in an inert carrier oil, and can serve as reaction vessels for biological assays. These drops can be manipulated with very high precision, with the carrier oil controlling the fluidics, ensuring the samples never contact the walls of the microfluidic channels. Small quantities of other reagents can be injected with a high degree of control. The drops can also encapsulate cells, enabling cell-based assays to be carried out. Hydrogel particles can be co-encapsulated within the drops to add genomic barcodes to the cell material, and thereby enable very high throughput sequencing with single-cell resolution. The use of these devices for studies of biology and for biotechnology applications will be described.

Both of these applications of drop-based microfluidics lead to new and interesting science, and to myriad applications, both in the production of new materials, and in biotechnology applications to improve human health.

# Merging centrifugal microfluidics with droplet microfluidics for rapid and multiplexed molecular diagnostics at the Point-of-Care

*Roland Zengerle - University of Freiburg*

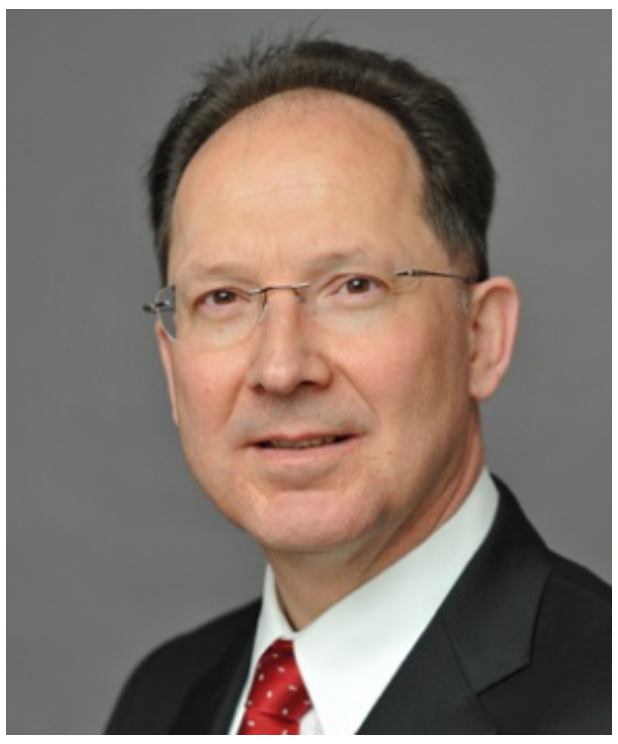

## Biography

*Full Professor with appointments at*

- *IMTEK - Department of Microsystems Engineering, University of Freiburg, Germany*
- *Hahn-Schickard, Villingen-Schwenningen & Freiburg, Germany*
- *BIOSS Centre for Biological Signalling Studies, Freiburg, Germany*

*Prof. Dr. Roland Zengerle is full professor at the Department of Microsystems Engineering at the University of Freiburg and director at the "Hahn-Schickard Institut für Mikroanalysesysteme". The research of Dr. Zengerle is focused on microfluidics and specialises in*

- *Point of care testing by centrifugal microfluidics*
- *Non-contact microdosage of picoliter & nanoliter volumes,*
- *Single cell printing and Bioprinting*
- *BioMEMS*

*Dr. Zengerle co-authored more than 300 papers. He is a member of the German national academy of sciences, Leopoldina.*

## Abstract

Centrifugal microfluidics enables efficient miniaturization, integration, parallelization and automation of biochemical assays especially for Point-of-Care applications. This talk will demonstrate fully integrated sample-to-answer genotyping assays from whole blood which enables rapid and multiplexed molecular diagnostics of infectious diseases. The assays are implemented on a LabDisk which is fabricated by micro-thermoforming from thin polymer films and are performed within 3 hours on a small and lightweight instrument.

Additionally advanced centrifugal unit operations will be presented such as centrifugal step emulsification merging droplet microfluidics with centrifugal microfluidics. This unit operation enables centrifugal droplet generation, DNA amplification and fluorescence detection within one single cavity of a disposable chip. It enables significant miniaturization of future devices for digital assays as well as acceleration of fully automated point of care testing within less than 1 hour. Results for absolute quantification of DNA by digital droplet-RPA,-LAMP and -PCR amplification using the novel unit operation are in good agreement with those obtained by a commercial state-of-the-art dPCR system. Finally, challenges and solutions for the scale-up of manufacturing the polymer LabDisk by microthermoforming will be discussed.